\documentclass[a4paper,12pt]{article}
\usepackage{eurosym}
\usepackage{amscd,amsfonts,amsmath,amssymb,amsthm,bbm,bm,latexsym,mathrsfs}
\usepackage{epsfig,graphics,graphicx,natbib,subfigure,longtable}
\usepackage[english]{babel}
\usepackage[usenames]{color,colortbl}
\usepackage{bookmark}
\usepackage{booktabs}
\usepackage{sgame}
\usepackage[flushleft]{threeparttable}
\usepackage{adjustbox}
\usepackage{sgame}
\usepackage[normalem]{ulem}
\usepackage{lineno} 
\usepackage[top=1in, bottom=1in, left=0.5in, right=0.5in]{geometry}
\usepackage{setspace}
\usepackage{ragged2e}
\usepackage{multirow}

\allowdisplaybreaks
\makeatother

\theoremstyle{remark}

\theoremstyle{plain}

\newcommand{\cg}{\cellcolor[gray]{0.95}}

\usepackage{tikz}
\usetikzlibrary{snakes}

\begin{document}

\title{\vspace{-50pt}  Are low frequency macroeconomic variables important for high frequency electricity prices?\thanks{This paper has previously circulated with the title ``Forecasting daily electricity prices with monthly macroeconomic variables''. We thank two anonymous referees, the associate editor and the editor for the useful comments. We thank seminar and conference participants at University of Geneva, University of Glasgow and Maastricht University, the 23rd Applied Time Series Econometrics Workshop at Federal Reserve Bank of St. Louis, the 13th Netherlands Econometric Study Group, the 2nd Workshop on Energy Economics at SKKU, the NBP Workshop on Forecasting at Central Bank of Poland, the 2nd IWEEE at Venice for helpful comments and suggestions to improve this work. The views expressed are those of the authors and do not necessarily reflect those of the European Central Bank (ECB).
This research used the Computational resources provided by INDACO Platform, which is a project of High Performance Computing at the University of Milan.
}
}
\author{
\hspace{-20pt}
Claudia Foroni\textsuperscript{a}
\hspace{15pt}
Francesco Ravazzolo\textsuperscript{b,c,d}
\hspace{15pt}
Luca Rossini\textsuperscript{e}\thanks{e-mail: \href{mailto: luca.rossini87@gmail.com}{luca.rossini87@gmail.com} -- \href{mailto: luca.rossini@unimi.it}{luca.rossini@unimi.it}}
\vspace{10pt}
 \\
        \small\textsuperscript{a}European Central Bank, Germany \hspace{5pt}
          \small\textsuperscript{b}BI Norwegian Business School, Norway\\
        \small\textsuperscript{c}Free University of Bozen-Bolzano, Italy\hspace{5pt}
          \small\textsuperscript{d}RCEA\hspace{5pt}
        \small\textsuperscript{e}University of Milan, Italy
  }

\date{\today}
\maketitle

\vspace{-1cm}
\abstract{

Recent research finds that forecasting electricity prices is very relevant. In many applications, it might be interesting to predict daily electricity prices by using their own lags or renewable energy sources. However, the recent turmoil of energy prices and the Russian-Ukrainian war increased attention in evaluating the relevance of industrial production and the Purchasing Managers' Index output survey in forecasting the daily electricity prices. We develop a Bayesian reverse unrestricted MIDAS model which accounts for the mismatch in frequency between the daily prices and the monthly macro variables in Germany and Italy. We find that the inclusion of macroeconomic low frequency variables is more important for short than medium term horizons by means of point and density measures. In particular, accuracy increases by combining hard and soft information, while using only surveys gives less accurate forecasts than using only industrial production data.


}

\clearpage


\section{Introduction}
\label{sec_Intro}

Electricity markets have received increased attention in the literature since their deregulation in the late 90s. Several reasons are motivating such interest. First, electricity is not storable, and therefore demand and supply must always match. To achieve this, sophisticated markets have been created, where the one-day-ahead hourly spot market is the main market in terms of volume. In the day-ahead spot market, hourly prices are set by matching demand and supply. This market offers a large amount of data and requires forecasts of both demand and prices. Second, power grids are one of the most critical infrastructures and have a major role in sustainable development and economic growth. The recent innovation in energy production and, in particular, the large increase in renewable energy resources (RES) have added complexity to the management of the electricity system, see \cite{Gianfreda2018, Gianfreda2022} for an application of RES to predict day-ahead prices. Moreover, smart grids are the future technologies in power grid development, management, and control, see \cite{YuCecati2011} and \cite{YuXue2016}. They have revolutionized the regime of existing power grids, by employing advanced monitoring, communication, and control technologies to provide a secure and reliable energy supply. And new technologies have changed energy consumption, making it necessary to use effective energy management strategies based on electricity prices and electricity load forecasts. As a consequence, a growing literature has investigated these dynamics and built forecasting models for several markets all around the world (Europe, United States, Canada, and Australia). Further, the recent turmoil caused in energy prices by the Russian invasion of Ukraine captured even more attention around the evolution of electricity prices.

The literature on load forecasting has focused on three horizons: short-term load forecasts (from one hour to one week); mid-term load forecasts (from one week to one month) and long-term load forecasts (from one month to years); see, for example, \cite{AlfaresNazeeruddin2002} and \cite{SuganthiSamuel2012} for definitions and models. By contrast, the literature on price forecasting has mainly focused only on the day-ahead spot market, see \cite{Weron2014} for a recent and detailed review. Two possible reasons are that the predictive power of predictors for day-ahead spot prices is usually short-lived, and longer future markets are subject to low liquidity and highly correlated to spot prices. This paper tries to fill this gap and introduces a new methodology to produce mid-term spot price forecasts, that is forecasts of day-ahead spot prices up to one month ahead. {We believe that storage-based system, such as storage hydropower plants, plans their production over a longer period to maximize profits and rely on mid-term probabilistic price predictions to take these decisions.} 

In order to accomplish this, in this paper we suggest to apply lower frequency predictors based on macroeconomic variables containing more valuable information for mid-term horizons as opposed to the regressors usually applied in short-term price forecasting. 
{ In particular, we focus on macroeconomic variables aiming at capturing the central role of industry and manufacturing in the use of energy. The goal of our paper is therefore to evaluate the relevance of industrial production and a forward-looking indicator of it, represented by the Purchasing Managers' Index output survey, in forecasting the daily electricity prices in the two countries with the largest share of manufacturing in the euro area: Germany and Italy. {Both countries are similar on the large dependency to oil and gas in the production of electricity as the recent invasion of Ukraine by Russia has underlined, and, as said above, the use of electricity is largely employed for the manufacturing sector, precisely up to 45\% in Germany of total electricity consumption and up to 40\% in Italy.\footnote{See, for example, \url{https://www.enerdata.net/estore/energy-market/germany/} for German statistics and \url{https://annuario.isprambiente.it/sysind/823} for Italian statistics.}} The use of the survey allows us to exploit its forward-looking component for improving the forecasts.}
Several papers have documented that surveys are useful for predicting macro variables, see e.g. \cite{hjl2005}, \cite{Abberger2007}, \cite{Claveria2007}, \cite{AJR2016}. As highlighted by, e.g., \citet{Evans05}, \citet{Giannone2008}, \citet{AGJT2013}, an advantage of surveys is that they are timely available and possibly contain forward-looking information. We label them ``soft'' data, given they usually just represent the opinions or impressions of consumers or purchasing managers, who are asked to compare economic and financial conditions today with the recent past, and/or to forecast the economic environment in the near future.\footnote{See blog \url{https://www.stlouisfed.org/on-the-economy/2017/may/hard-data-soft-data-forecasting}.} However, this sort of economic indicator surveys has not yet been compared to hard data in a context similar to ours, so the reliability in our context is not proven yet.

{To exploit the content in the hard and soft macroeconomic data we provide also an important methodological contribution: we develop a mixed frequency model which accounts for the mismatch in frequency between the daily prices and the monthly macro variables.} In the last years, there is a growing interest in models that account for data of different frequencies for forecasting purposes. The focus in the literature has mostly been on improving the forecast of low-frequency variables by means of high-frequency information. In particular, different models have been introduced for dealing with the different sampling frequencies at which macroeconomic and financial indicators are available. The most common choice is to reduce the model to state-space form and use the Kalman filter for forecasting
(e.g. see \cite{Aruoba2009, Giannone2008,Mariano2002} and in a  Bayesian context \cite{Eraker2015,Schorfheide2015}). As an alternative	choice, \cite{Ghysels2016} develops a class of mixed-frequency VAR model, where both low- and high-frequency variables are included in the vector of dependent variables \citep[see][for an application in small-scale factor model]{Blasques2016}. This class of model is estimated by OLS, but the number of regressors tends to increase due to the stacking structure of the model.

In a univariate context, \cite{Ghysels2006} introduce MIDAS, which links directly low- to high-frequency data \citep[see][for macroeconomic forecasting]{Clements2008,Clements2009}, but it requires a form of NLS estimation, which improves the computational costs substantially in a model with more than one high-frequency explanatory variables. \cite{Foroni2015} develop unrestricted MIDAS (U-MIDAS) model, which can be estimated by OLS and thus handle high-frequency explanatory variables. However, the U-MIDAS models have problems when the frequency mismatch is high and several regressors are included, thus leading to a Bayesian extension of the literature on MIDAS and U-MIDAS, see \cite{Foroni2015WP} and \cite{Pettenuzzo2016}, and a stochastic volatility estimation method for U-MIDAS in density nowcasting \citep{Carriero2015}.

Recently, new models have been proposed for forecasting high-frequency variables by means of low-frequency variables. An example is the paper of \cite{DalBianco2012}, which analyzes the forecasts of the euro-dollar exchange rate at the weekly frequency by means of macroeconomic fundamentals in a state-space form \`a la \cite{Mariano2009}. \cite{Ghysels2016} contributes by introducing a mixed-frequency VAR model, which address both the prediction of high-frequency variables using low-frequency variables and vice versa. Furthermore, \cite{Foroni2018} introduce Reverse Unrestricted MIDAS (RU-MIDAS) and  Reverse MIDAS (R-MIDAS) model for linking high-frequency dependent variable with low-frequency explanatory variables in a univariate context.

From a methodological innovation point of view, this paper proposes a Bayesian approach to RU-MIDAS of \cite{Foroni2018} in order to incorporate low-frequency information into models for the (probabilistic) prediction of high-frequency variables. Our goal is to derive a model that allows combining efficiently several predictors, possibly with different frequencies. The use of Bayesian inference allows to compute probabilistic statements without any further assumption. Despite other mixed-frequency specifications that could be incorporated, we decide to work with the RU-MIDAS because our focus is on longer horizons, where relative predictability is lower and linear models usually perform accurately.

{The empirical contribution of the paper stands in predicting the daily electricity prices at different horizons while including different low-frequency explanatory variables.} In the last years, a large and growing body of literature deals with the forecasting of daily electricity prices \citep[see][for a review]{Weron2014}. However, the main focus of the literature is on the forecasting of electricity prices influenced by variables with the same frequency, such as renewable energy sources \citep{Gianfreda2018, Gianfreda2022} or weather forecasts \citep{Huurman2012}. This empirical application draws on the literature using macroeconomic variables to improve the forecasting performance of single-frequency models, since macroeconomic variables are of interest in the diagnostic of electricity prices. 

The results show that there is a strong improvement in the forecasting if we add all monthly macroeconomic
variables (such as PMI surveys and IPIs), different oil prices specification (daily or monthly) {or natural gas price}, at almost all horizons for Italy and at the short horizons for Germany. We find gains around 20\% at short horizons and around 8\% at long horizons. 
{We compare the results against different benchmark models, where we include various dummy specifications, and the inclusion of monthly macroeconomic variables leads to better forecasting results at all the horizons.}
Interesting, accuracy increases by combining hard and soft information, and using only surveys give marginally less accurate forecasts than using only industrial production data. {This is of a particular relevance nowadays, in which the macroeconomic activity and the prospects of a downturn caused by the Russia-Ukraine war can turn out to be particularly useful in understanding and predicting the movements in the electricity prices.} 

The paper is organized as follows. Section \ref{sec_Model} summarizes the RU-MIDAS models and the Bayesian approach. Section \ref{sec_Data} presents the data used in the paper. In Section \ref{sec_Results}, we present the forecasting of daily electricity prices by using daily and monthly macroeconomic variables. Section \ref{sec_Conclusions} concludes.

%
%
\section{RU-MIDAS model}
\label{sec_Model}

\cite{Foroni2018} show the derivation of the reverse unrestricted MIDAS (RU-MIDAS) regression approach from a general dynamic linear model and its estimation procedure.
Here we sketch the derivation, adapting it to our case of monthly/daily observations. 
If we assume only daily observations, the model considered is an autoregressive model (AR), where the lags of the daily explanatory variables are included. Once we include a less frequent series, such as a monthly time series, each day we consider again an AR model extended with the less frequent series up to the last data point available. This last data point will not change for 28 days (4 weeks), but the distance between it and the daily observations changes day by day. In order to take into account these changes we include a set of dummy variables, such that we do not lose any information.

For the sake of simplicity, we assume the following two variables of interest. Let us observe at high-frequency (HF) the variable $x$ for $t=0,\frac{1}{k}, \ldots, \frac{k-1}{k},1,\ldots$, while the variable $y$ can be observed at low frequency (LF) every $k$ periods for $t=0,1,2,\ldots$.

In our case, the variable $x$ follows an AR($p$) process

\begin{equation}
c(L) x_t = {f(L)} y_t^{\ast} + e_{xt}, \label{AR_spec}
\end{equation}
where $y^{\ast}$ is the exogenous {unobserved regressor sampled at higher frequency}; {$f(L) = f_1L+\ldots+f_pL^p$, $c(L) = I-c_1L-\ldots - c_pL^p$} and the errors are white noise. Furthermore, we assume that the starting values $y^{\ast}_{-p/k},\ldots,y^{\ast}_{-1/k}$ and $x_{-p/k},\ldots,x_{-1/k}$ are all fixed and equal to zero.

It is possible to introduce the lag operator for the low and high-frequency variables. In particular, let us define $Z$, the LF lag operator such that $Z = L^k$ and $Z^jy_t = y_{t-j}$; and the polynomial in the HF lag operator, $\gamma_0(L)$ with $\gamma_0(L){f(L)}$ containing only $L^k=Z$. If we multiply Eq. \eqref{AR_spec} by $\gamma_0(L)$ and $\omega(L)$, we have
\begin{equation}
\gamma_0(L) c(L) \omega(L)x_t = \gamma_0(L){f(L)}\omega(L) y_t^{\ast} + \gamma_0(L) \omega(L) e_{xt}, \,\quad\, t=0,1,2,\ldots \label{mul_AR_spec}
\end{equation}
where $\omega(L) = \omega_0 + \omega_1L + \ldots + \omega_{k-1}L^{k-1}$ represents the temporal aggregation scheme by means of a polynomial. Moreover, if Eq. \eqref{mul_AR_spec} is represented as
\begin{equation}
\tilde{c}_0(L) x_t = g_0(Z)y_t + \tilde{\gamma}_0(L) e_{xt}, \,\quad\, t=0,1,2,\ldots, \label{exact_RUMIDAS}
\end{equation}
where $y_t=w(L)y_t^{\ast}$ and $g_0(Z)$ is the product of $\gamma_0(L)$ and ${f(L)}$ and function only of $Z$, Eq. \eqref{exact_RUMIDAS} is called an exact reverse unrestricted MIDAS model.
In particular, in Eq. \eqref{exact_RUMIDAS}, the high-frequency variable is a function of its own lags, of the LF lags of the observable variable $y$ and of the error terms. Thus, the HF period influences the model specification. For each $i=0,\ldots,{k-1}$, a lag polynomial in the HF lag operator, $\gamma_i(L)$, can be defined and the product $g_i(L) = \gamma_i(L){f(L)}$ is a function only of power of $Z$. As seen above, if we multiple Eq. \eqref{AR_spec} by $\gamma_i(L)$ and ${f(L)}$, we have
\begin{equation}
\tilde{c}_i(L)x_t = g_i(L^{k+i})y_t + \tilde{\gamma}_i(L)e_{xt},\, \quad \, t=0+\frac{i}{k},1+\frac{i}{k}\ldots, \, \quad \, i=0,\ldots,k-1 \label{RUMIDAS}
\end{equation}
such that a period structure in the RU-MIDAS is introduced.

Since the parameters of Eq. \eqref{AR_spec} are unknown and also $\gamma_i(L)$ cannot be determined exactly, it is possible to use an approximate reverse unrestricted MIDAS (RU-MIDAS) models based on linear lag polynomial
\begin{equation}
\tilde{a}_i(L)x_t = b_i(L^{k+i})y_t + \xi_{it}, \, \quad \, t=0+\frac{i}{k},1+\frac{i}{k}\ldots, \, \quad \, i=0,\ldots,k-1 \label{approx_RUMIDAS}
\end{equation}
where the orders of $\tilde{a}_i(L)$ and $b_i(L^{k+i})$ are larger enough such that $\xi_{it}$ is a white noise. Since the error terms $\xi_{it}$ are correlated across $i$, one could estimate the RU-MIDAS equations for different values of $i$ by using a system estimation method. In particular, Eq. \eqref{approx_RUMIDAS} can be grouped in a single equation by adding a proper set of dummy variables.  {In our empirical application, we consider a daily dependent variable and monthly explanatory variables, and we use 28 days (4 weeks) for modeling purpose. The specification of 28 days (4 weeks) lags is used to include seasonality and the low frequency macroeconomic variables.
}

{Let $x_t=0$ and $y_t=0$ for $t\le 0$ as the initial condition. Let $d=1,\ldots,28$ and define $x_t$ the high frequency variables for $t=\frac{1}{28},\frac{2}{28},\ldots$. As a further step, we define $D_d(t)=1$ for $t=\frac{d}{28},1 + \frac{d}{28},2 + \frac{d}{28},\ldots$, thus for $t= n + d/28$, where $n=0,1,2,\ldots$, $D_d \cdot y_{t-\frac{d}{28}} = y_n$ and $D_{d'} \cdot y_{t-\frac{d'}{28}} = 0$ for $d \ne d'$. Then }
the single-equation version of Eq. \eqref{approx_RUMIDAS} can be written as
{\begin{align}
x_t & = \sum_{d=1}^{28} \alpha_d D_d y_{t-\frac{d}{28}}
+ \sum_{d=1}^{28} \beta_{1,d} D_d x_{t-\frac{1}{28}} + \sum_{d=1}^{28} \beta_{2,d} D_d x_{t-\frac{2}{28}} + \sum_{d=1}^{28} \beta_{3,d} D_d x_{t-\frac{7}{28}} + v_t \, \quad \, t= \frac{1}{28},\frac{2}{28},\ldots,  \label{RUMIDAS}
\end{align}}
where $D_1,D_2,\ldots,D_{27}, D_{28}$ are dummy variables taking value one in each second day, third day, $\ldots$, 28-th day and the first day of the month respectively.

{
As an example,  Figure~\ref{Dummy_Daily} provides the product between the daily
representation of the Industrial Production Index (IPI) for Consumer goods in Germany
and the dummy variables for the first dummy observed for the whole sample (left panel) and zoomed on the period from 01/01/2006 to 31/12/2010 (right panel).
}

{\[
\left[\text{Insert Figure \ref{Dummy_Daily} here}\right]
\]}

In conclusion,  the error term, $v_t$, is independent and identically distributed as a Normal distribution, where $\mathcal{N}(0,\sigma^2)$.
{Regarding the daily dependent variable, $x_t$, we consider lag $1,2$ and $7$, meaning the previous day, two days before and one week before the delivery time as suggested by \cite{Weron2008}, \cite{Raviv2015} and \cite{Gianfreda2018}.}
It is possible to estimate the model in Eq. \eqref{RUMIDAS} by GLS to allow the possible correlation and heteroskedasticity. However, it may be difficult to estimate the model by using a frequentist approach, in particular if there are several regressors. Thus we use a Bayesian approach to solve this issue.

\subsection{Bayesian approach}

Contrary to most of the MIDAS literature, which follows a classic approach, in this paper we estimate our models with Bayesian techniques. Few papers so far have focused on the Bayesian estimation of regular MIDAS models (see, for example, \cite{Pettenuzzo2016} and \cite{Foroni2015WP}). However, the Bayesian method has not yet been applied to the RU-MIDAS approach, as described in the previous section. Differently than the classical estimation, our Bayesian approach allows for estimation of complex nonlinear models with many parameters, is useful for imposing parameter restrictions and, above all, allows to compute probabilistic statements without any further assumption.

In this paper, therefore, we focus on introducing the Bayesian estimation in the RU-MIDAS model. 
We define prior information on the vector of coefficients and on the variance, using the independent Normal-Wishart prior as in \cite{Koop2010} adapted to univariate time series, thus a Normal-Gamma prior.
This section is devoted to the study of prior and posterior inference on the vector of coefficients of the autoregressive model and on the variance coefficient. In particular, we work with a prior which has AR coefficients and variance coefficients being independent each other, thus it is called independent Normal-Gamma prior.

The general prior for this kind of model, which does not involve the restrictions of the natural conjugate prior, is the independent Normal Gamma prior. Let us assume $\gamma$ be the vector of the AR coefficients defined in equation \eqref{RUMIDAS} and made by $\alpha_1,\dots,\alpha_{28},\beta_{1,1},\dots, \beta_{1,28},\ldots,\beta_{3,1},\dots, \beta_{3,28}$ and $\sigma^2$ be the variance coefficients, thus the independent prior can be represented as $p(\gamma,\sigma^{-2}) = p(\gamma)p(\sigma^{-2})$. In this case, the prior for $\gamma$ is a normal distribution:
\begin{equation}
\gamma \sim \mathcal{N}\left(\underline{\gamma},\underline{V}_{\gamma}\right), \label{Gamma_Prior}
\end{equation}
while the prior for the variance coefficient is a Gamma distribution
\begin{equation}
\sigma^{-2} \sim \mathcal{G}a(\underline{a},\underline{b}) \label{Sigma_Prior}
\end{equation}
By using these priors, the joint posterior $p(\gamma,\sigma^{-2}|x)$ has not a convenient form, but the conditional posterior distribution have a closed form. In particular, the posterior distribution for the vector of AR coefficients is:
\begin{equation}
\gamma|x,\sigma^{-2} \sim \mathcal{N}\left(\overline{\gamma},\overline{V}_{\gamma}\right) \label{Gamma_Posterior}
\end{equation}
where the posterior mean and posterior variance are:
\begin{align*}
\overline{V}_{\gamma} &= \left(\underline{V}_{\gamma}^{-1} + \frac{1}{\sigma^2} \sum_{t=1}^T z_t z_t\right)^{-1}\\
\overline{\gamma} &=  \overline{V}_{\gamma}\left(\overline{V}_{\gamma}\underline{\gamma} + \frac{1}{\sigma^2} \sum_{t=1}^T z_t x_t\right),
\end{align*}
where $z_t$ is the vector containing the explanatory variables $y_{t-\frac{1}{28}}, \dots, y_{t-\frac{28}{28}}$ and the lagged dependent variables $x_{t-\frac{1}{28}},\dots,x_{t-\frac{7}{28}}$.

Moreover, the posterior distribution for the variance coefficients is:
\begin{equation}
\sigma^{-2}|\gamma,x \sim \mathcal{G}a(\overline{a}, \overline{b}) \label{Sigma_Posterior}
\end{equation}
where the posterior hyperparameters are
\begin{align*}
\overline{a} &= \frac{T+\underline{a}}{2} \\
\overline{b} &= \underline{b} + \sum_{t=1}^T (x_t - z_t \gamma)^2
\end{align*}

We estimate the Bayesian model described above using Markov chain Monte Carlo (MCMC) methods. {In detail, we {draw} from the posterior distributions in Eq.~\eqref{Gamma_Posterior} and \eqref{Sigma_Posterior} by Gibbs sampling and }
 all our results are based on samples of 6.000 posterior draws, with a burn-in period of 1.000 iterations. Moreover, we choose the prior hyperparameters such that the priors are not informative\footnote{{For the vector of AR coefficients and monthly macroeconomic variables, we set the prior mean, $\underline{\gamma}$, and the prior variance, $\underline{V_{\gamma}}$,  equal to $0$ and a matrix with diagonal elements equal to $10$, respectively. Regarding the Gamma prior for the variance coefficient, we adopt $\underline{a} = \underline{b} = 0.1$, such that $E[\cdot] = 1$ and $Var(\cdot) = 10$.}}.

{Regarding the forecasting techniques adopted in the paper, we use the direct forecasting method \citep[see, e.g.][]{Marcellino2006} since the forecasting of the future values of the explanatory variable $y$ is not required, although the model specification should change for each forecasting horizon considered. The long forecast horizons we target, up to 28 days ahead, might imply that short-term dynamics are less relevant and we investigate in section \ref{sec_Results} different lag specifications.}

%
\section{Data Description}
\label{sec_Data}

In this section, we describe the two datasets analyzed in the application. In particular, we consider {two of the largest} European countries from a macroeconomic and energy point of view, Germany and Italy, both parts of the G8 economies.

We use daily day-ahead prices (in levels and averaged across the twenty-four hourly day-ahead prices) to estimate models for electricity traded/sold in Germany and Italy. Moreover, we employ different monthly macroeconomic variables, which either differ by country, such as industrial production index or Purchasing Managers' Index (PMI); or are equal for all the countries, such as the oil brent prices. The national electricity prices are obtained directly from the corresponding power exchanges. In particular, the German daily auction prices of the power spot market is collected from the \textit{European Energy Exchange} {EPEX}, whereas the daily single national prices PUN are collected from the Italian ISO.

In terms of macroeconomic variables, we consider the total industrial production index for Germany and Italy, and its main components: consumer goods (IPI-Cons, i.e. the consumer durable goods); electricity (IPI-Elec, i.e. the activity of providing electric power, natural gas, steam, hot water and the like through a permanent infrastructure (network) of lines, mains, and pipes) and manufacturing (IPI-Manuf, i.e. the activities in the manufacturing section involve the transformation of materials into new products). The data are taken from Eurostat and are seasonally and calendar adjusted.

The other macroeconomic variable consider is the Manufacturing PMI surveys, which is a
measure of the performance of the manufacturing sectors and it is derived from a survey of $500$ industrial companies.
On the other hand, the Italian PMI is based on surveys of about $400$ industrial companies.
The Manufacturing PMI is based on five industrial indexes with the following weights: New Orders ($30\%$), Output ($25\%$), Employment ($20\%$), Suppliers' Delivery Times ($15\%$), and Stock of Items Purchased ($10\%$) with the Delivery Times Index inverted so that it moves in a comparable direction. This index is of capital importance since the manufacturing sector dominates a large part of the total GDP and thus it is an important indicator of the business conditions and of the overall economic condition in the country.
Moreover, a reading above $50$ indicates an expansion of the manufacturing sector compared to the previous month; on the other hand, a value below $50$ represents a contraction. We consider it an important soft macroeconomic indicator of future economic conditions.

We should emphasize that the monthly variables are released on different days of the month and thus we need to keep attention when we will analyse them in the forecasting exercise. As an example, the Industrial production index is released on the first working day of the following month, thus the IPI for November 2018 is released on the 2nd of January 2019; the IPI for January 2019 is released on the 1st of March, etc. Obviously, if the first day of the month is a specific holiday or Saturday or Sunday, then the IPI will be released on the first coming working day. For what concerns the surveys, the final release of the PMI is typically available in the early morning of the first day of the month after the one they refer to (i.e. the PMI of December is released at the very beginning of January, etc.) {as explained in Fig.~\ref{fig:time_line}.}

{\[
\left[\text{Insert Figure \ref{fig:time_line} here}\right]
\]}

The sample spans from 1 January 2006 to 31 December 2019 for both countries. We use the first seven years as an estimation sample and the last seven years as a forecast evaluation period. The historical dynamics of these series are reported in Fig. \ref{Graph_Ger} for Germany and in Fig. \ref{Graph_Ita} for Italy. Prices clearly show the new stylized fact of ``downside'' spikes together with mean-reversion. {On the other hand, Fig. \ref{Graph_Oil_Gas} shows the daily and monthly dynamics of the oil prices jointly with the daily gas prices. The black (daily) and the red (monthly) lines show the same course during the entire sample size.} In particular, the oil price shows two strong falls, the first around the end of 2008 and the beginning of 2009; the second around the end of 2014. Regarding the first fall, the drop in oil prices that started in 2008 takes place against the backdrop of the global financial crisis. In fact, the oil prices drop from historic highs of $141.06\$$ in July 2008 to $40.07\$$ in March 2009. After an increase in the oil prices in the following years, the second fall appears in the fourth quarter of 2014 as robust global production exceeded demand, thus leading to a sharp decline.

{
\[
\left[\text{Insert Figure \ref{Graph_Oil_Gas} here}\right]
\]
}

\[
\left[\text{Insert Figure \ref{Graph_Ger} here}\right]
\]

Regarding the other macroeconomic variable of interest, the industrial production index (IPI), it shows a different behavior between the two countries. In fact, in Germany, the industrial production index follows the first drop of the oil prices in 2008/2009, while it leads to a constant slow increase in the following years until 2018. In recent years (2018 and 2019), the IPI related to the different specifications has a slow decrease, with a particularly strong fall for the IPI of electricity supply in 2019. On the other hand, the situation in Italy is completely different since, after the fall in 2008, the situation remains the same or slightly decreases in the subsequent years, with a tiny increase at the end of 2017.
Regarding the PMI Survey, the behavior of the series follows the same movement across the two countries, with a huge fall in 2008, which represents an important contraction in both countries economies and a consequent increase in 2009 and a constant behavior in the next years over $50$, which indicates a small expansion of the economies. As stated for the IPI, the last years of the sample show evidence of contraction in both countries and in both economies as can be seen for the GDP.

\[
\left[\text{Insert Figure \ref{Graph_Ita} here}\right]
\]

%
\section{Empirical Results}
\label{sec_Results}

In this section, we present the results for the forecasting of daily electricity prices by means of different macroeconomic variables. In particular, the first estimation sample in the forecasting exercise extends from January 2006 to December 2012, and it is then extended recursively by keeping the size of the estimation window fixed to $7$ years 
in such a way we perform a rolling window estimation. For each day of the evaluation sample, we compute forecasts from $1$ to $28$ days ahead, and we assess the goodness of our forecasts using different point and density metrics.

\subsection{Forecasting framework}

Regarding the accuracy of point forecasts, we use the root mean square errors (RMSEs) for each of the daily prices and for each horizon. Whereas, to evaluate density forecasts, we use both the average log predictive score, viewed as the broadest measure of density accuracy \citep[see][]{Geweke2010} and the average continuous ranked probability score (CRPS). The latter measure does a better job of rewarding values from the predictive density that are close and not equal to the outcome, thus it is less sensitive to outlier outcome \citep[see, e.g.][]{Gneiting2007,Gneiting2011}.

As seen in Eq. \eqref{RUMIDAS}, one can evaluate different RU-MIDAS model based on different lags order of the high-frequency variables and on the inclusion of different low-frequency variables.
As suggested in \cite{Weron2008}, \cite{Raviv2015} and \cite{Gianfreda2018}, we consider a RU-MIDAS model with lag order of the electricity prices equal to $7$. In particular, this model includes only the first, second, and the seventh lag of the daily electricity prices; with abuse of notation, we will call the model a {sparse AR(7)} model. Moreover, due to the seasonal components of the daily electricity prices, we include seasonal dummies representing each season of the year: spring, summer, autumn, and winter, respectively. In the benchmark models, called BAR(3), the estimation is provided by using a Normal-Gamma prior and the same prior has been used also for the Bayesian RU-MIDAS model, called B-RU-MIDAS\footnote{{The models are estimated using Matlab 2020b on a Macbook Pro with an Intel Core i7 @2.70 GHz processor and 16 GB memory. The computational times to obtain $6.000$ posterior draws under each model specification on the full sample size is changing from 0.40 seconds to 3.85 seconds depending on the number of variables included.}}.

{In order to check the good performance of our proposed model, we have included different benchmark models in the analysis. In particular, we have included day-of-the-week dummies \citep{Ziel2018}, which take values $1$ for the $k-$th day of the week, where $k$ is equal to $1$ for Monday and $7$ for Sunday. For this specific choice of dummies, we have used four different benchmark models: a BAR with $7$ lags (based on the BIC choice), a BAR with $1$ lag, a {sparse BAR with $7$ lags}, and a BAR with $4$ lags (as suggested by the auto.arima \citep{Makridakis2020} function on the full sample size). }{{The model benchmark equation could be represented as
\begin{align*}
x_t = \sum_{l=1}^{p_{\text{max}}} \phi_l x_{t-l}  + \sum_{j=1}^7 \psi_j \text{DoW}_d^j + v_t, \, \text{ where } \, \text{DoW}_d^j = \begin{cases} 1, \text{ if $d$ is the $j$-th day of the week} \\ 0, \text{ otherwise } \end{cases}
\end{align*}
where $\text{DoW}_d^j$ means the day-of-the-week dummy for $j=1$ (Monday) until $j=7$ (Sunday) and $p_{\text{max}}$ is the maximum lag chosen by the four methods previously described.
}}

{As a robustness check, we have also run the forecasting exercise without including the seasonal dummies in all the models, but we do not find any interesting results\footnote{These results are not presented in the paper due to lack of space and are available in the Supplementary Material.}. Moreover, we focus our analysis on the RU-MIDAS model when only one lag of the daily electricity prices is included\footnote{These results are not presented in the paper due to lack of space and are available in the Supplementary Material.}.}

The main interest of the paper is forecasting daily electricity prices by using macroeconomic variables. Thus, we consider different macroeconomic explanatory variables in the construction of the models. In each model and for each country, as explanatory variables, we include separately the monthly specification of the Manufacturing PMI surveys or the three main industrial production indices (All-IPI). As further check, we add a specification of the model that has both all the three main industrial production indices and the PMI surveys. Moreover, the daily oil prices\footnote{The daily oil price has been interpolated over the weekends in order to have a full sample size.} and the monthly oil prices has been included in the model specification for some specific cases. {In order to check the influence of the oil price, we have included a model, where we include the daily natural gas price with the three main IPI and the PMI surveys and a model with only daily natural gas price.}
When we discuss the three main IPI, we consider IPI based on the manufacturing sector (IPI-Manuf), on the activity of providing electric power (IPI-Elec) and on Main Industrial Groupings (MIG) for consumer goods (IPI-Cons). As a further check, we analyse models where we include PMI surveys and either ALL-IPI, only one of the index, or combinations of two indices (IPI-Cons-Elec, IPI-Cons-Manuf, IPI-Elect-Manuf) and we include or not the oil price specification as monthly or daily {and first differences of the IPI}\footnote{Due to lack of space, these results are not presented in the paper and are available upon request.}.

{In detail, in our tables, we report the RMSE, average log predictive score and average CRPS for all the models with seasonal dummies or with day-of-the-week dummies. Since we do not consider a single benchmark model, but we consider a group of benchmark models, the ones with day-of-the-week dummies and the BAR with seasonal dummies, we have employed the Model Confidence Set procedure of \cite{hansen_etal.2011} to jointly compare the predictive power of all models. We use the \textsf{R} package \textsf{MCS} detailed in \cite{bernardi_catania.2016} and differences are tested separately for each class of models (meaning for each panel in the tables and for each horizon) with a confidence level of $\alpha=0.1$.}

\subsection{Forecasting Results}\label{results}

\subsubsection*{Point forecasts}
We start by evaluating the point forecast of the different models and in the panel (A) of Table \ref{Ger_Forec_AR3_Seas_6models} and \ref{Ita_Forec_AR3_Seas_6models}, we present the RMSEs for different mixed frequency models relative to the group of benchmark models, the so called Bayesian {sparse BAR(7)} with seasonal dummies and Normal-Gamma prior; BAR(7), {sparse BAR(7)}, BAR(1), and BAR(4) with day-of-the-week dummies.

\[
\left[\text{Insert Table \ref{Ger_Forec_AR3_Seas_6models} here}\right]
\]

Focusing first on Germany, in Table \ref{Ger_Forec_AR3_Seas_6models} we observe that the RMSE remains broadly constant over the horizons. Since we are predicting daily electricity prices, there is a strong improvement in the forecasting if we add monthly macroeconomic variables. In particular, the improvement is large in the first horizons, and in general for short-term forecasts, while at longer horizons, such that $21$ and $28$, the content of macro information is less relevant and we even see a decrease in the forecasting performance, even if gains are still 10\% relative to the benchmark. It is in general hard to rank the models with different macroeconomic indicators, where the performance of the different model specifications in terms of point forecasting is rather similar. In particular, adding oil price with daily or monthly specification does not improve the forecasting results with respect to the model that does not include oil price. {The inclusion of daily natural gas price in the RU-MIDAS specification leads to small improvement with respect to the models that include the oil prices.} As further results, the inclusion in the analysis of only one of the two most important macroeconomic variables, PMI surveys or All-IPI, leads to the worst performance with respect to models that consider both variables, and using only surveys gives less accurate forecasts than using only industrial production data.

However, what we find, is strong evidence of statistically superior predictability by the alternative models to the benchmark at several horizons. The B-RU-MIDAS model with All-IPI, PMI surveys, and oil price gives the best statistic at one day ahead with a $20\%$ reduction in RMSE with respect to the class of benchmark models, but also another version of B-RU-MIDAS without including the oil price provides economically sizeable gains at those horizons.
Moreover, B-RU-MIDAS with all the IPI variables and PMI surveys provide also {gains statistically significant} at longer horizons, such that $h=21,\,28$ with no differences between the inclusion or not of the oil price (both daily or monthly).
{{Looking at Germany, the model that includes all the macroeconomic variables and the monthly oil prices is considered the best model at the first two horizons, while increase the horizons lead to different results in terms of model confidence set. For example at $21$ step ahead, the best model becomes the one that includes only the PMI surveys or only all the Industrial Production Indexes, while at $28$ step ahead, the model that includes all the macroeconomic variables and the oil prices return to be the best model. In terms of the best RMSEs across the models, the B-RU-MIDAS with all the macroeconomic variables and the daily oil price is the best across the horizons, followed by the one with monthly oil price and daily natural gas price.}

\[
\left[\text{Insert Table \ref{Ita_Forec_AR3_Seas_6models} here}\right]
\]

{For the case of Italy, results are shown in Table \ref{Ita_Forec_AR3_Seas_6models}. Contrary to the case of Germany, for the case of Italy, there is a strong movement of the RMSEs from the first horizon to the $28$ horizon, moving from $8.11$ to $10.64$ for the {sparse BAR(7)} with seasonal dummies or from $6.66$ to $10.5$ for the BAR(7) with day-of-the-week dummies. Moreover,  the model that considers All-IPI, PMI Surveys, and the daily oil prices analyzed in the paper tend to dominate in terms of forecasting performance. In particular, the B-RU-MIDAS with all the IPI macroeconomic variables and the daily oil prices leads to a reduction of around $19\%$ of the RMSE with respect to class of benchmark models at the first horizon. While at the second and third horizon, the gains are in terms of $21\%$. On the other hand, when the horizon size increases, the B-RU-MIDAS models gain somewhat less, but still, the reduction is around $9\%$ from the class of benchmark models. Looking at the best model across the horizons, we can see that the RU-MIDAS with daily oil price and all the macroeconomic variables is the best model, followed by the model with daily natural gas price, and the one with only all the macroeconomic variables in the first horizons, and with monthly oil price in the last horizons.}  
Differently from Germany, in Italy, the model that includes all the macroeconomic variables and the monthly oil price is not considered the best model across the horizon from the evidence of statistically superior predictability in terms of Model Confidence Set. Whilst the best model across the horizons seems to be the model that includes only all the IPIs (All-IPI) or all the IPIs and the PMI surveys jointly. Moreover, in Italy from a statistically superior predictability point of view, the inclusion of oil price, in terms of both monthly or daily specification, does not {lead} to better models at all the horizons.

{As a robustness check, we have compared the models with and without seasonal dummies. The first result is that the inclusion of the seasonal dummies leads to improvements in point forecasting at all the horizons and for all the different model specifications. These results are valid for both countries and all the horizons analyzed. On the other hand, we have conducted a second robustness check, where we have included in the RU-MIDAS specification only one lag of the electricity prices. For this scenario, we have analyzed a scenario with seasonal or without seasonal dummies. In Germany, the inclusion of only one lag of electricity prices leads to the worst results with respect to the inclusion of three lags. However, the inclusion of one lag improves the forecasting performance with respect to the class of benchmark models and the model with seasonal dummies, all the macroeconomic variables and the daily oil price results in the best model across all the horizons except for the first horizon, where the best model is the one that includes the daily natural gas price. On the other hand, in Italy, the inclusion of one lag of the electricity prices leads to small improvements relative to the first $3$ horizons. For these horizons, the best model results in the BAR model with seven lags and without macroeconomic variables, while at the longer horizons, the best model returns the one with macroeconomic variables and daily oil prices. }

All in all, we can conclude that in terms of point forecasting, the inclusion of macroeconomic variables, such as the industrial production index and the PMI surveys jointly with the oil price, is very helpful in predicting electricity prices in Germany and Italy.

\subsubsection*{Density forecasts}

We now focus on two different metrics for the density forecasts: the CRPS and the log predictive score, the Panel (B) and Panel (C) in Table \ref{Ger_Forec_AR3_Seas_6models} and \ref{Ita_Forec_AR3_Seas_6models}, for Germany and Italy respectively. In general, the accuracy of density forecasts improves in the models with macroeconomic variables, where we observe substantial low CRPS across the models and horizons. As before, we observe generally higher CRPS values when the horizon increases from $1$ to $28$. {In particular, as in the point forecast analysis, the B-RU-MIDAS with all the monthly IPI variables (All-IPI); PMI Surveys, and daily gas price specification gives the best statistics at one day ahead with a $23\%$ reduction in average CRPS in Germany
Secondly, when both oil price specification (daily or monthly) are includes in the model leads to improvements against the class of benchmark models.
 At longer horizons, as in the point forecast analysis, the inclusion of macro information and oil price leads to lower gains, but still significant and around $8\%$ better relative to the benchmark models. Moreover, the inclusion of only one monthly macroeconomic variable (such as All-IPI or PMI Surveys) separately leads to small improvements in the order of $19\%$ at short horizons and $3\%$ at long horizons. Across long horizons, we can see that the inclusion of all the macroeconomic variables plus daily oil or monhtly oil or daily gas prices improves the density forecasting measures. As for point forecasting, using only surveys gives less accurate forecasts than using only industrial production data. For the inclusion in the Superior set of Models at $10\%$, at the first horizon, the model with all monthly macroeconomic variables and daily natural gas price is the best, while from the second horizon to the $14$ horizons, the models with oil or gas specifications are the best. The only change comes at the $14$ and $21$ horizons, when the models with only one macroeconomic variables are strongly included in the Superior set of models. In the last horizon, the situation comes back to the initial situation with the best models the one with oil or gas prices and macroeconomic variables.}

{Looking at Panel (B) of Table \ref{Ita_Forec_AR3_Seas_6models}, we can  see that the model with daily oil prices and monthly macroeconomic variables outperforms the class of benchmark models at all the horizons with bigger improvements at longer horizons. As stated for Germany and in the point forecasting exercise, the inclusion of only one macroeconomic variables leads to improvements from a Superior set of models, while the model with oil or gas specifications are never included in the best set.}

{{Regarding the average log predictive score (see Panel (C) in Table \ref{Ger_Forec_AR3_Seas_6models} and \ref{Ita_Forec_AR3_Seas_6models}), the results change with respect to the average CRPS.  In particular, for Germany, the average log predictive likelihood shows smaller increases at all horizons except for the first and last horizons. The gains in term of log predictive score is higher in the models that include only the monthly macroeconomics variables jointly moving from a $11\%$ at the first horizons to a $9\%$ at the last horizons. In this case, there are no evidences of superior predictability of a model over the others, except that the models that include monthly macroeconomic variables and monthly oil prices leads to gains with respect to the benchmark models. In particular, as short horizons (second, third and seventh), the best model seems to be the one with BAR(7) with day-of-the-week dummies, although the model included in the superior set are the one with gas or oil prices.}

{For Italy, the gains of using macroeconomic variables is more stable over the horizons. In particular across the horizons with some exception, the best model is the one with all macroeconomic variables and daily gas prices followed by the one with daily oil prices. Regarding the inclusion in the Superior set, the models with only all the industrial production indexes and the one with them and the PMI are always included across the horizons, while the one with daily oil are including in the medium horizons (third, seventh and fourteenth).
 Also if we include the oil price specification in term of monthly or daily price leads to improvements of the forecasting accuracy over the horizons but at lower quantity.} 

{As further robustness check, in the Supplementary Material, we have included in the analysis different specification of the models without seasonal dummies and with only one lag of electricity prices. Regarding the first specification, as in the point forecasting analysis, there are improvements when in the analysis are included seasonal dummies with respect to the models without them. These results are obtained for the two density forecasting measures analyzed.}

{Regarding the inclusion of one lag of electricity prices in the RU-MIDAS specification, in Germany, the models always beat the class of benchmark models across the horizons for the average CRPS. These results are partially confirmed for the average log predictive likelihood, where for the second and third horizon, the best model is the benchmark model with $7$ lags and seasonal dummies. In Italy, for the average CRPS, the improvements is evident for long horizons, in fact at short horizon the best models are the one with seven lags and day-of-the-week dummies. The situation changes completely at long horizons, when the inclusion of macroeconomic variables and daily oil prices leads to better and strong results. In conclusion, for the average log predictive likelihood in Italy, the inclusion of macroeconomic variables with or without seasonal dummies leads to better model specification.}

%
\section{Conclusions}
\label{sec_Conclusions}

This paper analyses for the first time to the best of our knowledge the forecasting performances of mixed frequency models for electricity prices. In particular, we use monthly macroeconomic variables for predicting daily electricity prices in {two of the largest} European countries, Germany and Italy. The paper studies how to incorporate low-frequency information from manufacturing Purchasing Managers' Index surveys and industrial production index into models that forecasts high-frequency variables, the daily electricity prices. {Moreover, in the analysis, we have included different specifications of the oil prices, measured as the monthly or daily variables, and the daily natural gas prices.}

{Our analysis of point and density forecasting performances covers different horizons (from one day to one month ahead) on the sample spanning from 2013 to 2019. Our results clearly indicate that the RU-MIDAS specifications with all monthly macroeconomic variables and the inclusion of oil prices or natural gas price dominate different class of benchmark models, both in terms of point and density forecasting over all the horizons. {This result is very relevant to understand electricity price dynamics, in the period going from the COVID pandemic to the most recent Russian invasion of Ukraine.} {Moreover, for Germany, we find gains around $20\%$ at short horizons and around $8\%$ at long horizons against standard benchmark models with seasonal dummies and around $8\%$ at all the horizons against the strong benchmark model (the BAR(7) with day-of-the-week dummies). For Italy, these gains have a similar magnitude with respect to the standard benchmark model, while they reduced to $4\%$ against the best model among the benchmarks}. Thus it turns out that the macroeconomic low-frequency variables are more important for short horizons than for longer horizons. Moreover, the benchmark model is rarely included in the model confidence set. }

We conclude that from an energy forecasting perspective these mixed frequency models seem to have interesting and important advantages over simpler models. Going forward, it would be interesting to study the possible extension of these models to hourly data in order to include other variables of interest, such as renewable energy sources, which are currently taking lead in the electricity generation, { and forecasted demand. However,
electricity demand depends on temperature, and accurately forecasting weather for several weeks ahead is extremely difficult.}  {The line of research presented in this paper can be extended by proposing a model linking the current novel non-linear literature (such as random forests, deep neural networks or additive regression trees) and the RU-MIDAS formulation.}
%
%
\bibliographystyle{apalike}
\bibliography{Midas_Biblio}

%
\clearpage
\appendix{}

\begin{figure}[h!]
\centering
\caption{{\bf{Dummy indexing representation.}}}
\vspace{-0.1in}
\begin{justify}
\footnotesize{{Series for Industrial Production Index (IPI) for Consumer Goods in Germany by using the first dummy variable observed from 01/01/2006 to 31/12/2019 (left) and from 01/01/2006 to 31/12/2010 (right). }}
\end{justify}
\begin{tabular}{cc}
\includegraphics[width=8cm]{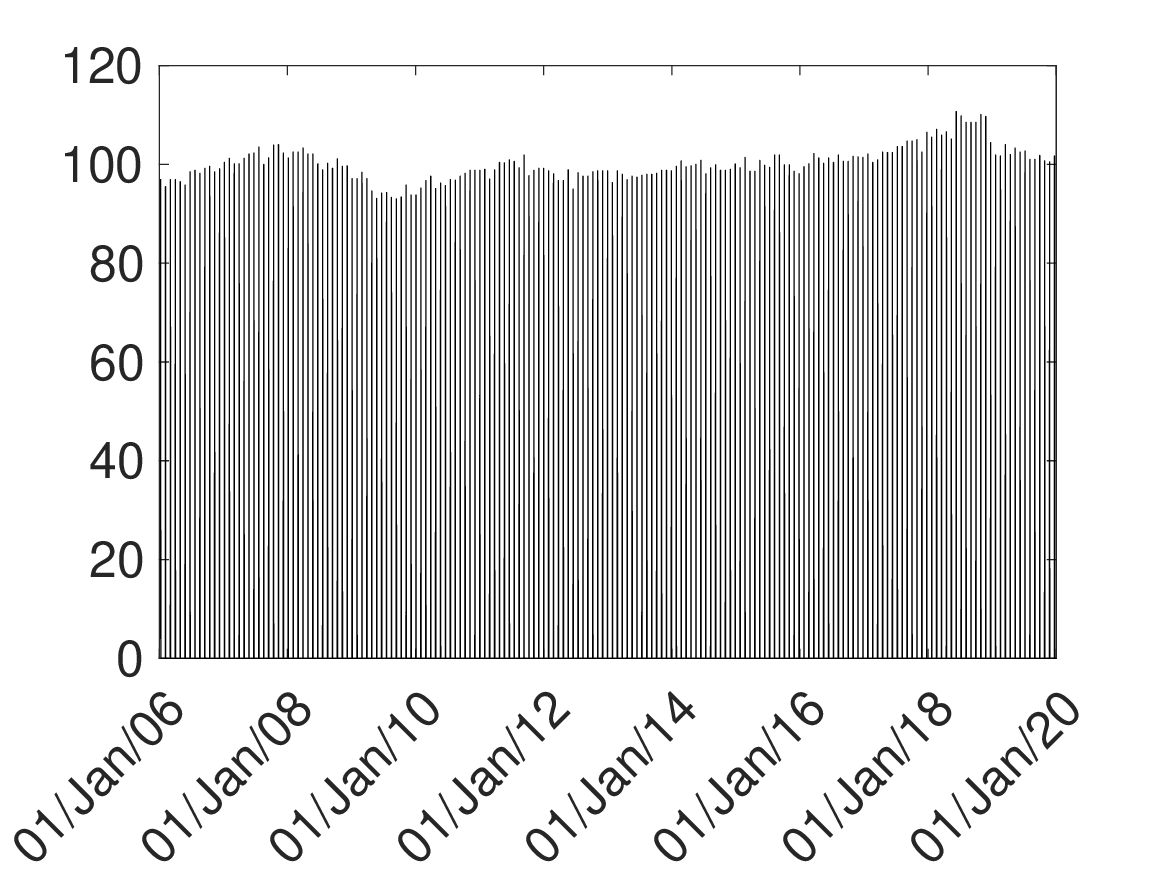} &
\includegraphics[width=8cm]{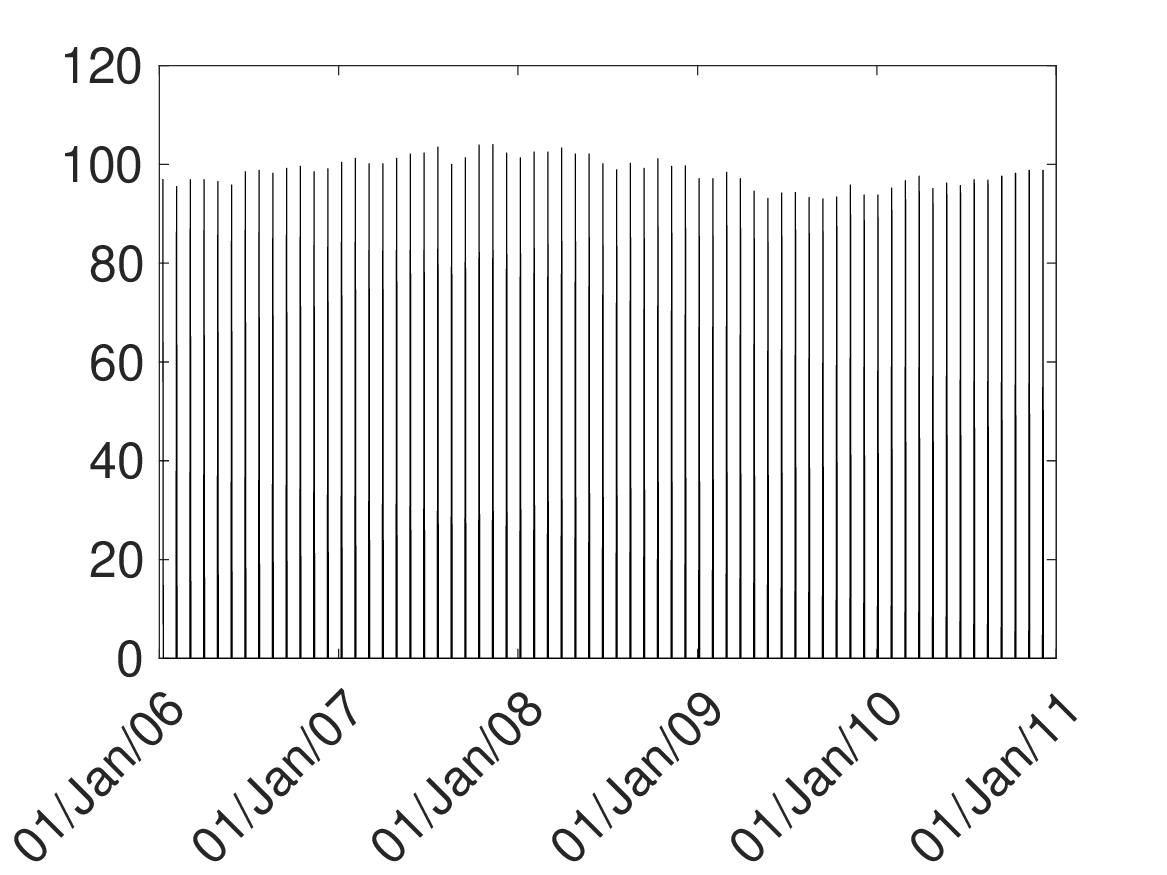} \\
\end{tabular}
\label{Dummy_Daily}
\end{figure}

{\begin{figure}[h!]
\centering
\caption{\bf{Example of Time line of release date for different macroeconomic variables.}}
\vspace{0.1in}
\resizebox{\linewidth}{!}{
\begin{tikzpicture}[]
\draw (0,0) -- (15/1.10,0);
\foreach \x in {0, 8, 15}{
   \draw (\x/1.10,3pt) -- (\x/1.10,-3pt);
}
\draw (0,0) node[below=3pt] { Released Industrial Production } node[below=14pt]{  Index for November 2018 } node[above=3pt] { January 2 2019  };
\draw (8/1.10,0) node[below=3pt] { Released PMI Survey } node[below=14pt]{ Flash for January 2019 } node[above=3pt] { January 24 2019  };
\draw (15/1.10,0) node[below=3pt] { Released PMI Survey} node[below=14pt]{ Final for January 2019 } node[above=3pt] { February 1 2019  };
\end{tikzpicture}
}
\label{fig:time_line}
\end{figure}
}

\begin{figure}[h!]
\centering
\caption{{\bf{Oil and Gas Data Representation}}}
\vspace{-0.1in}
\begin{justify}
\footnotesize{{Series for Daily (black) and Monthly (red) Oil Brent Prices (left) and for Daily Gas prices observed from 01/01/2016 to 31/12/2019.}}
\end{justify}
\begin{tabular}{cc}
\includegraphics[width=8cm]{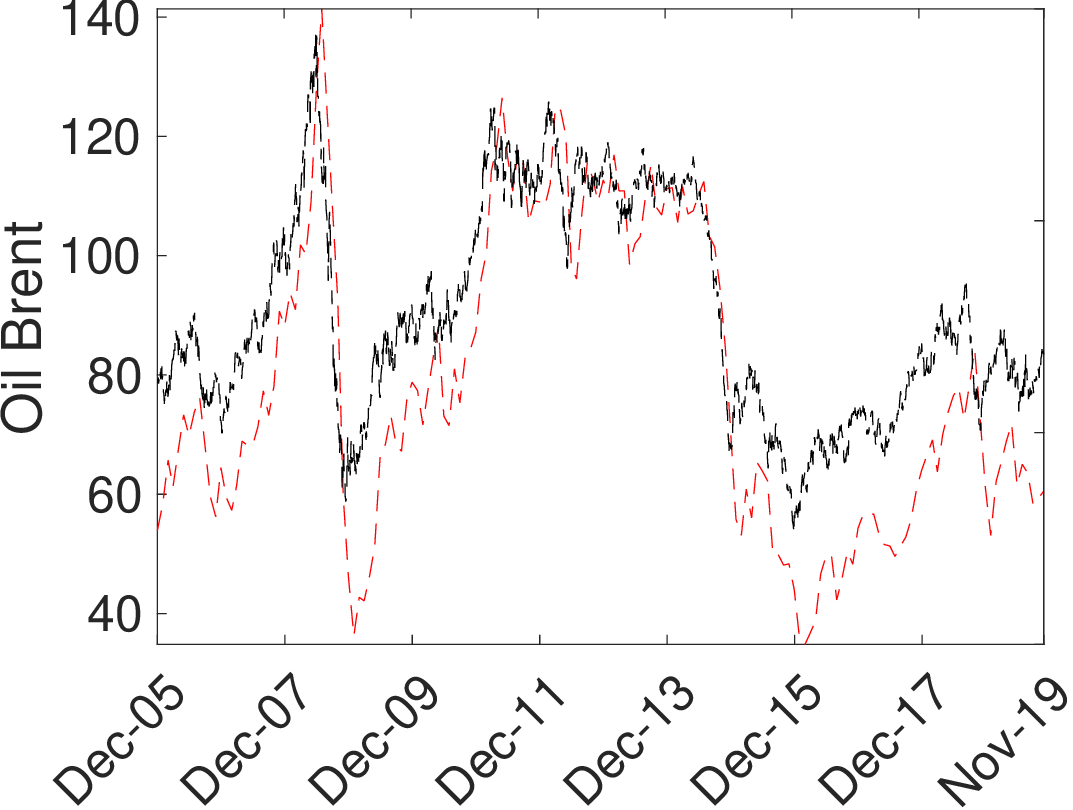} &
\includegraphics[width=8cm]{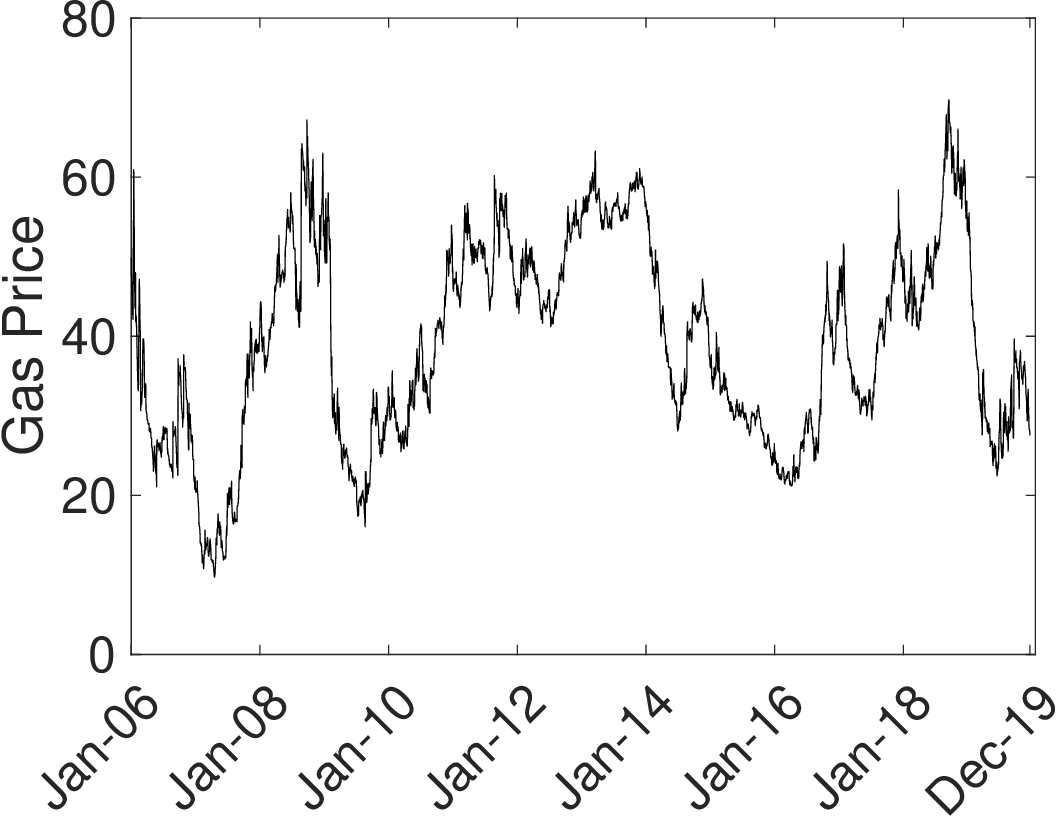} \\
\end{tabular}
\label{Graph_Oil_Gas}
\end{figure}

\begin{figure}[h!]
\centering
\caption{\bf{Germany Data Representation}}
\vspace{-0.1in}
\begin{justify}
\footnotesize{Daily Series for Electricity Day-ahead Prices (top left), Monthly PMI Surveys (top right), Monthly Industrial Production index (IPI) for Consumer Goods (middle left), Monthly IPI for Electricity Prices (middle right) and Monthly IPI for Manufacturing (bottom center) observed in Germany from 01/01/2016 to 31/12/2019.}
\end{justify}
\begin{tabular}{cc}
\includegraphics[width=8cm]{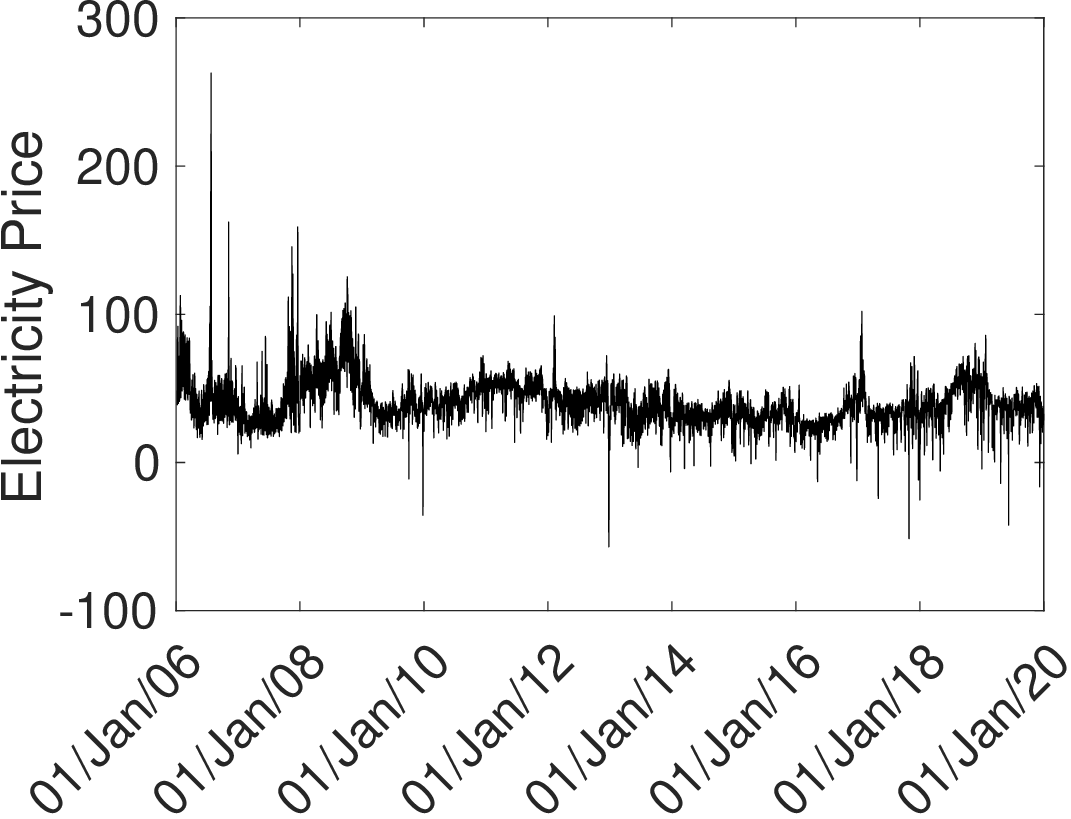} &
\includegraphics[width=8cm]{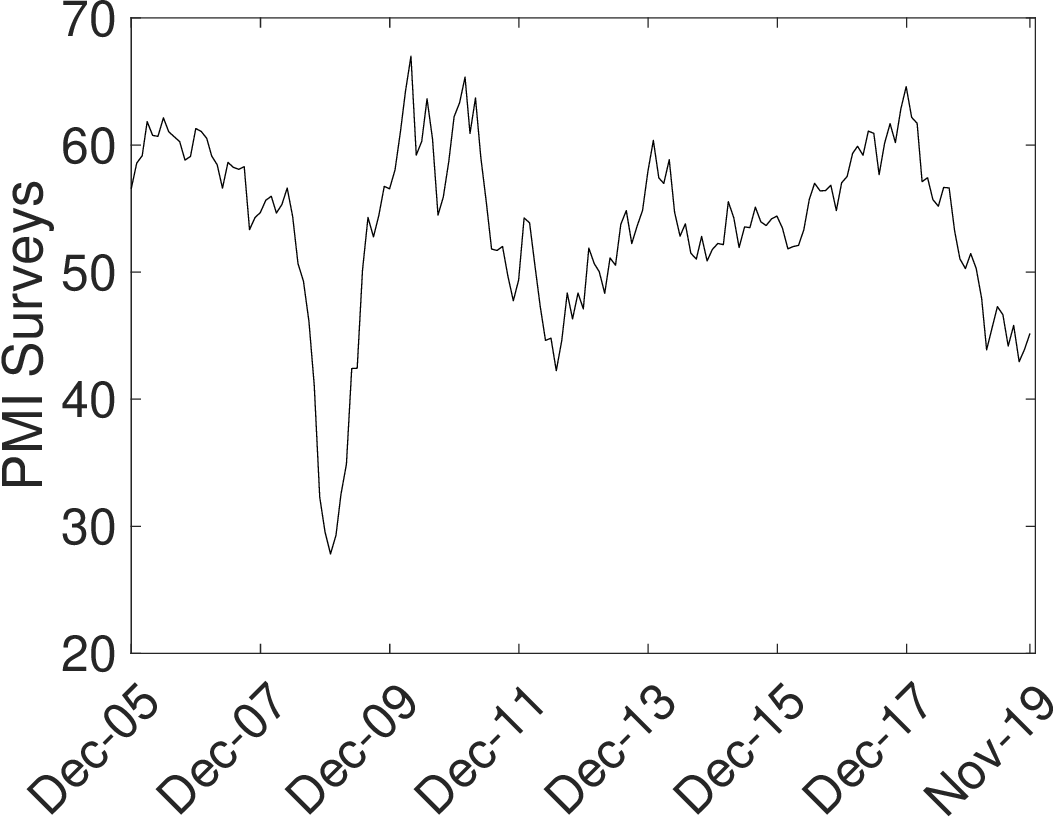} \\
\includegraphics[width=8cm]{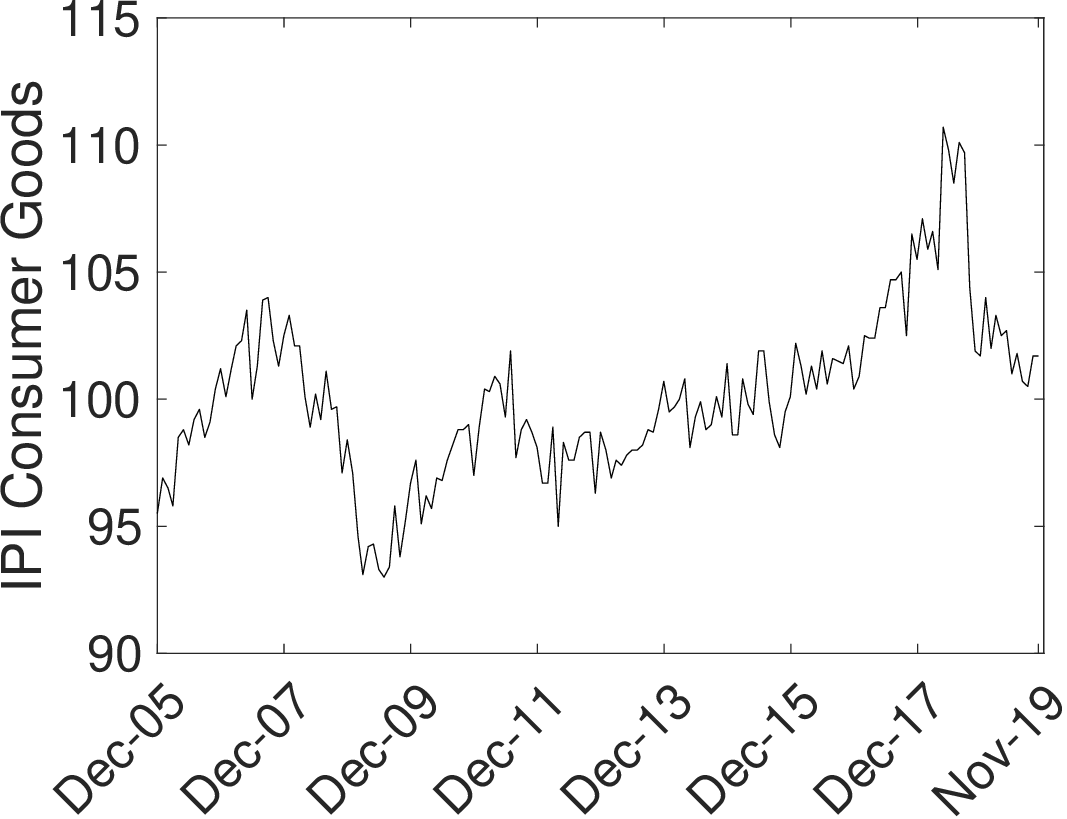} &
\includegraphics[width=8cm]{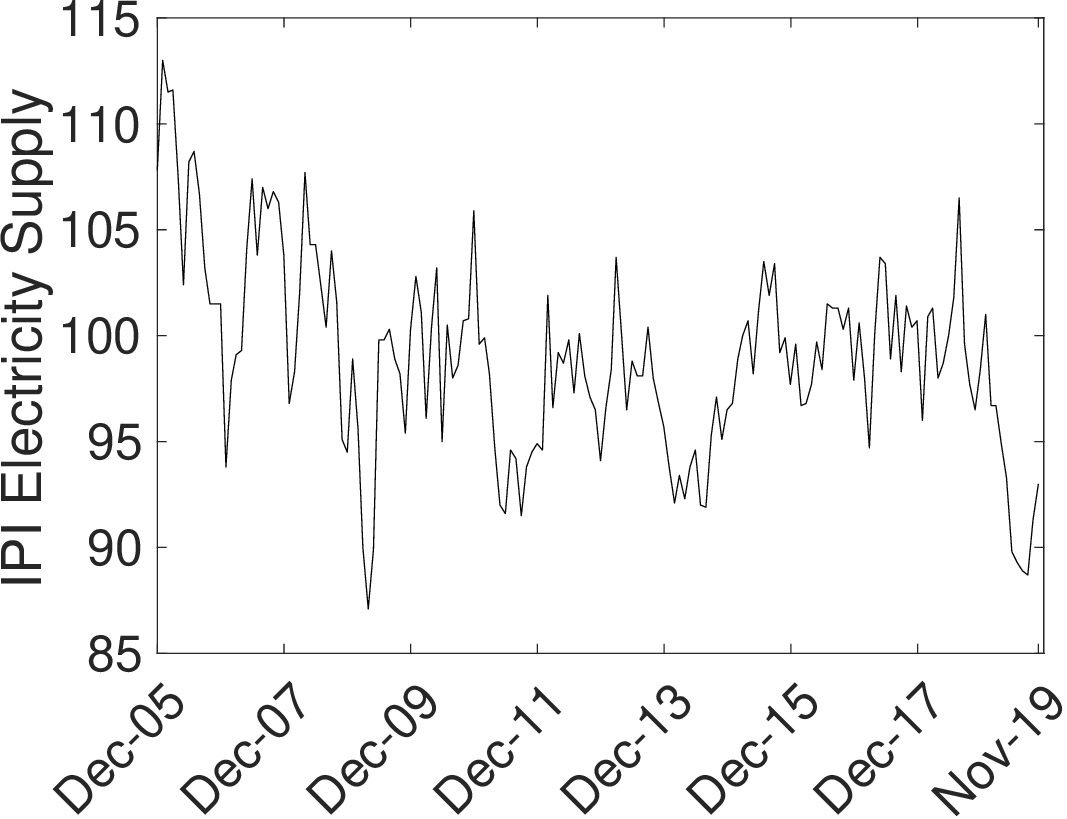} \\
\multicolumn{2}{c}{\includegraphics[width=8cm]{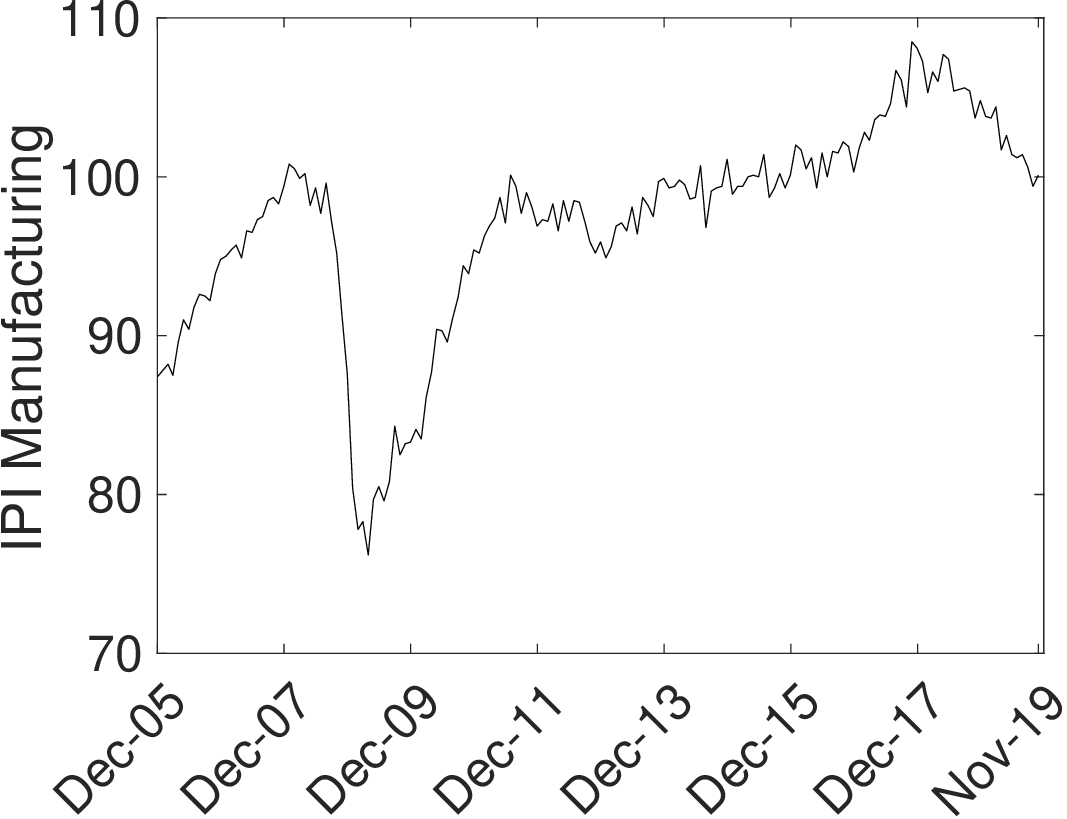}}
\end{tabular}
\label{Graph_Ger}
\end{figure}

\begin{figure}[h]
\centering
\caption{\bf{Italy Data Representation}}
\vspace{-0.1in}
\begin{justify}
\footnotesize{Daily Series for Electricity Day-ahead Prices (top left), Monthly PMI Surveys (top right), Monthly Industrial Production index (IPI) for Consumer Goods (middle left), Monthly IPI for Electricity Prices (middle right) and Monthly IPI for Manufacturing (bottom center) observed in Italy from 01/01/2016 to 31/12/2019.}
\end{justify}
\begin{tabular}{cc}
\includegraphics[width=8cm]{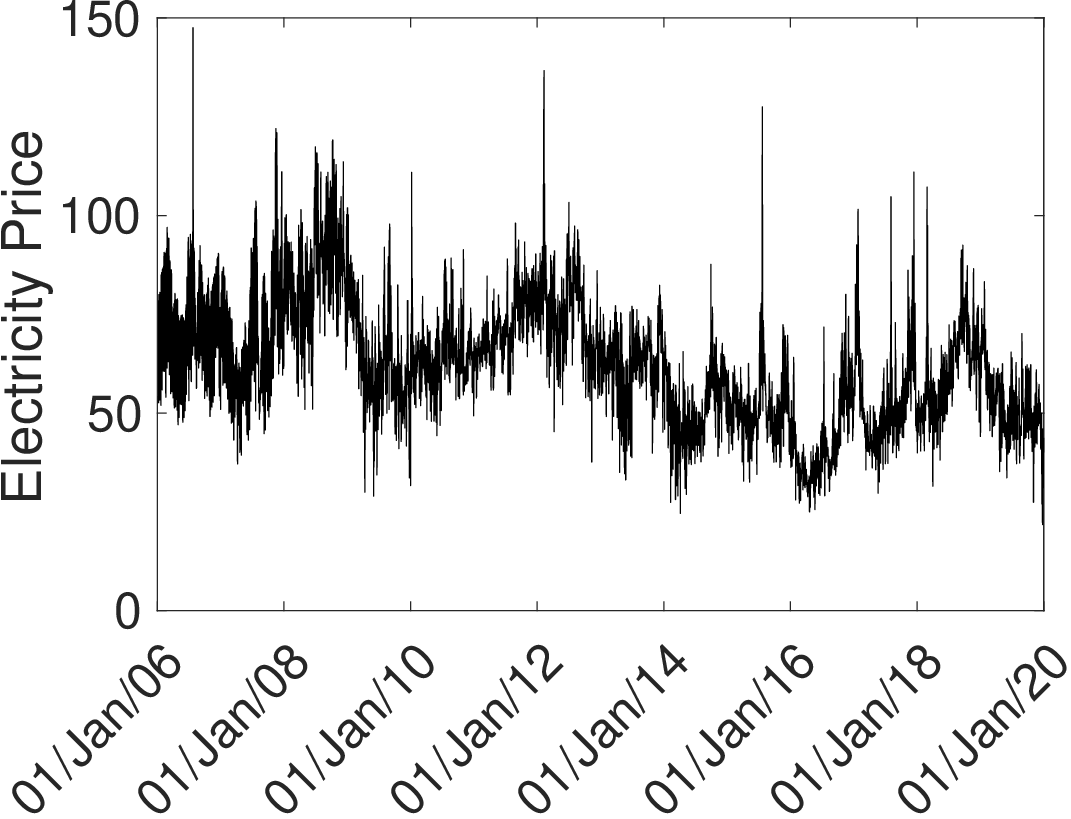} &
\includegraphics[width=8cm]{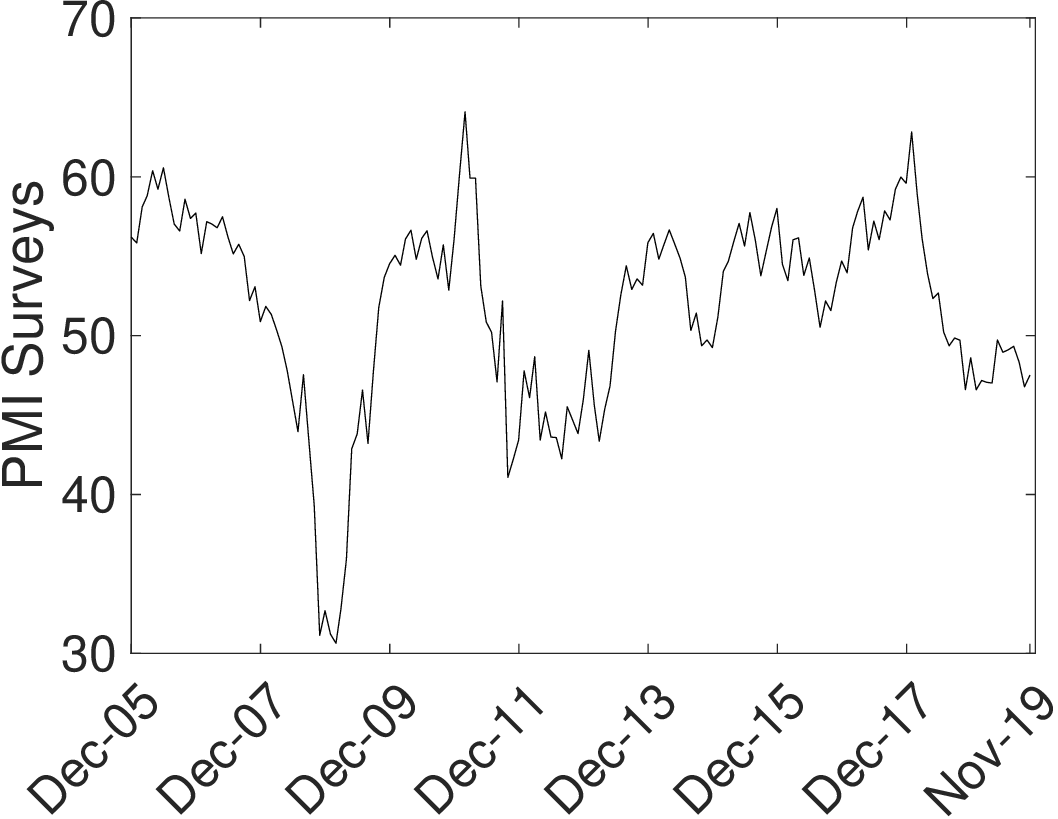} \\
\includegraphics[width=8cm]{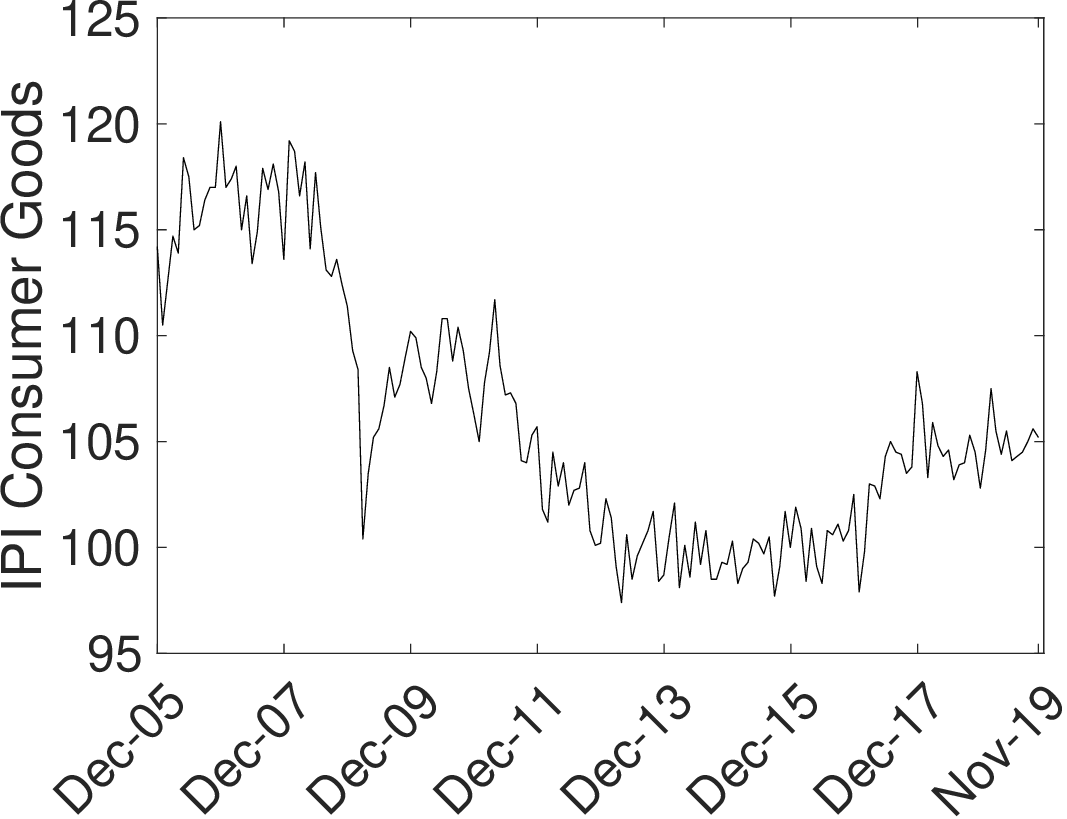} &
\includegraphics[width=8cm]{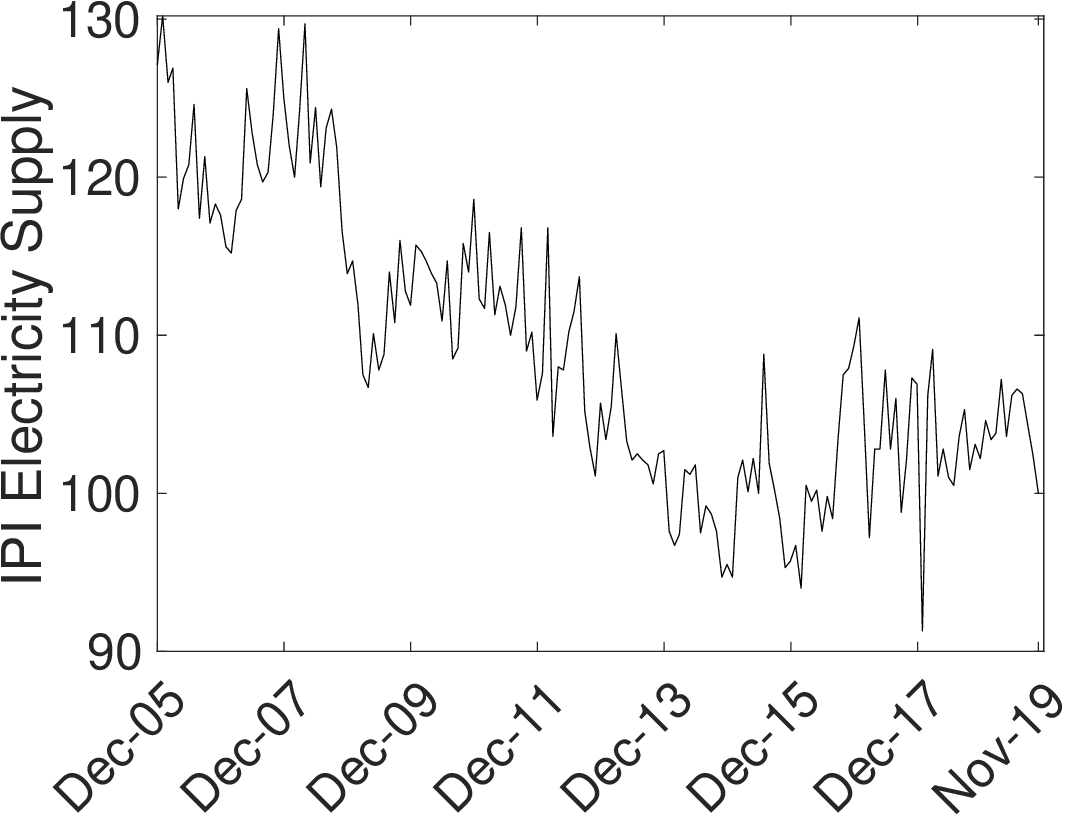} \\
\multicolumn{2}{c}{\includegraphics[width=8cm]{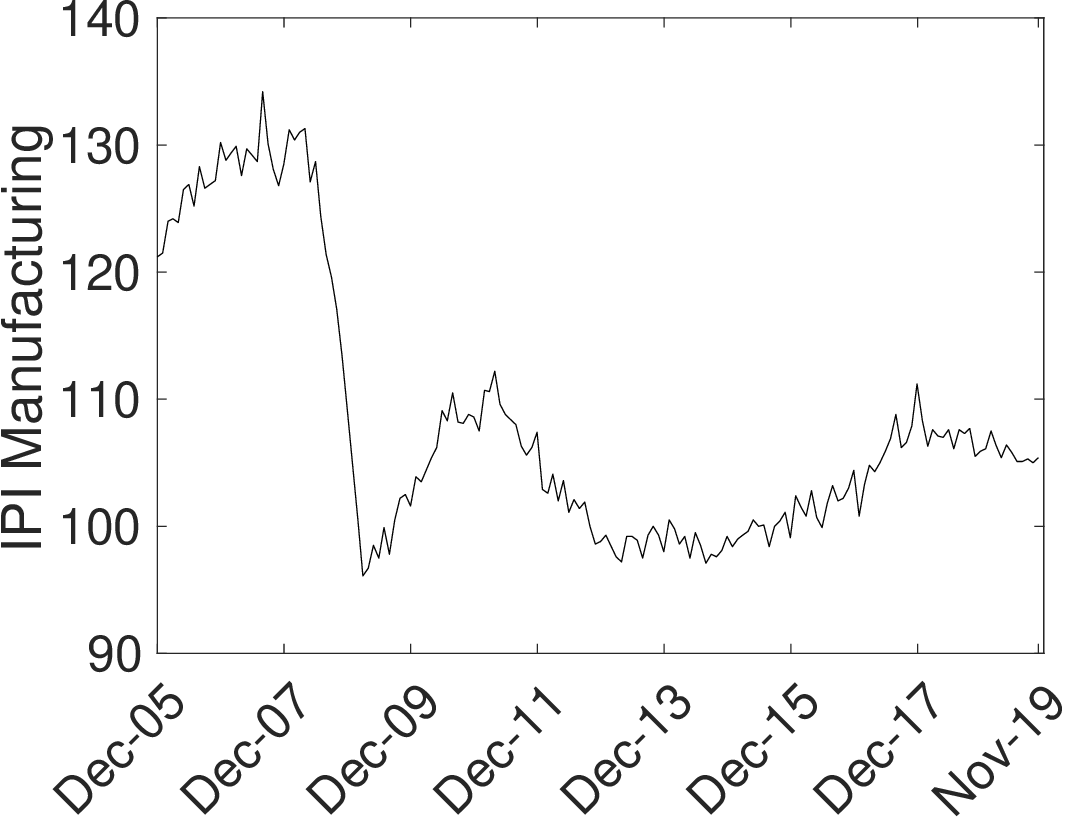}}
\end{tabular}
\label{Graph_Ita}
\end{figure}

\clearpage

\begin{table}[h!]
\caption{\bf Point and density forecasting measures  for Germany.}
\centering
\vspace{-0.1in}
\begin{justify}
\footnotesize{RMSE (Panel A); average CRPS (Panel B) and average Predictive Likelihood (PL) (Panel C) for all the models with day-of-the-Week or seasonal dummies. The estimation and forecasting sample last 7 years each. The forecasting is provided for horizons $h = 1, 2, 3, 7, 14, 21, 28$.}
\end{justify}
\begin{adjustbox}{width=\textwidth,center=\textwidth}
\begin{threeparttable}
\begin{tabular}{c|lccccccc}
\hline \\[-0.4cm]
Panel A (RMSE) & Horizon & 1 & 2 & 3 & 7 & 14 & 21 & 28 \\[0.01mm]
\hline \\[-0.4cm]
& BAR(7) &  9.758 &   10.248 &   10.502 &    10.846 & \cg  11.308 & \cg   11.637 &    11.951\\
\multirow{1}{*}{{{\textbf{with Day-of-the-Week}}}} & {sparse BAR(7)}  & 10.483  & 11.586 &  12.016 &   10.989 &    11.485 &    11.889 &   12.120 \\
\multirow{1}{*}{{{\textbf{Dummies}}}} & BAR(1)  &    12.068  &   13.102  &   13.538  &  \cg 12.730  &  13.398  &  13.601  &  14.237 \\
& BAR(4) (auto.arima) &    10.835 &    11.068 &     10.670 &   \cg   11.832 &     12.364 &   12.508 &    13.240 \\
\hline\\[-0.4cm]
& {sparse BAR(7)} & 11.359 & 12.546 & 12.903 & 11.154 & 11.588 &  11.977 & 12.167 \\
& {sparse BAR(7) (Daily Gas)} &   { 11.068} &   {12.086} & {12.365} &  {10.929} & {  11.308} &  {11.662} &   {11.726}\\
\multirow{4}{*}{{{\textbf{with Seasonal}}}}& B-RU-MIDAS (All-IPI + Surveys) & 9.127 & 9.737 & 10.121 & 10.412 & 10.832 & 11.104 & 11.341 \\
\multirow{4}{*}{{{\textbf{Dummies}}}}& B-RU-MIDAS (Surveys) & 9.101 & 9.717 & 10.107 & 10.386 & \cg 10.838 & \cg 11.087 & 11.361 \\
& B-RU-MIDAS (All-IPI) & 9.117 & 9.718 & 10.105 & 10.405 & \cg 10.835 &\cg  11.092 & 11.327 \\
& B-RU-MIDAS (All-IPI + Surveys + Monthly Oil) & 9.081 &\cg  9.660 & \bf 10.020 & 10.332 & 10.767 & 11.047 & \cg 11.304 \\
& B-RU-MIDAS (All-IPI + Surveys + Daily Oil) & 9.069 &\cg  \bf 9.655 & 10.021 & \bf 10.321 & \bf 10.741 & \bf 10.998 & \cg \bf 11.225 \\
& B-RU-MIDAS (All-IPI + Surveys + Daily Gas)  & \cg \bf 9.067 &  \cg 9.674 & \cg  10.045 &\cg  10.364 &    10.773 &  11.055 & \cg 11.251 \\ \\[-0.2cm]
\hline
\hline \\[-0.4cm]
Panel B (average CRPS) & Horizon & 1 & 2 & 3 & 7 & 14 & 21 & 28 \\[0.01mm]
\hline \\[-0.4cm]
& BAR(7) &      5.133 &  5.396 &   5.532 &      5.738 &     5.995 &     \cg  6.200 &      6.363\\
\multirow{1}{*}{{{\textbf{with Day-of-the-Week}}}} &  {sparse BAR(7)} &     5.605 &   6.233 &     6.422 &     5.834  &     6.085 &   6.321 &  6.447 \\
 \multirow{1}{*}{{{\textbf{Dummies}}}} &BAR(1) &     6.417   &  6.985  &    7.237  &   6.731  &     7.167  &      7.254  &      7.560 \\
& BAR(4) (auto.arima) &      5.764 &     5.855 & 5.619 &    6.302 &    6.687 &  6.737 &  7.088 \\
\hline\\[-0.4cm]
& {sparse BAR(7)} & 6.154 & 6.834 & 6.979 & 5.933 & 6.142 & 6.369 & 6.472 \\
& {sparse BAR(7) (Daily Gas)}   &     {5.961} &     {6.505} &  {6.632} &    {5.798} &    {5.984} &   {6.204}  &  {6.256}  \\
\multirow{4}{*}{{{\textbf{with Seasonal}}}} & B-RU-MIDAS (All-IPI + Surveys) & 4.785 & 5.144 & 5.333 & 5.509 & \cg 5.753 & 5.920 & 6.065 \\
\multirow{4}{*}{{{\textbf{Dummies}}}}&B-RU-MIDAS (Surveys) & 4.784 & 5.152 & 5.347 & 5.512 & \cg 5.774 & \cg 5.929 & 6.082 \\
& B-RU-MIDAS (All-IPI) & 4.784 & 5.144 & 5.332 & 5.514 & \cg 5.762 & \cg 5.925 & 6.067 \\
& B-RU-MIDAS (All-IPI + Surveys + Monthly Oil) & 4.746 & \cg \bf  5.079 & \cg \bf 5.258 &\cg  5.462 & 5.704 & 5.873 & \cg 6.039 \\
& B-RU-MIDAS (All-IPI + Surveys + Daily Oil) & 4.759 & \cg 5.102 & \cg 5.282 &\cg  5.461 & 5.687 & \bf 5.846 & \cg 5.999 \\
& B-RU-MIDAS (All-IPI + Surveys + Daily Gas) &  \cg  \bf  4.739   & \cg 5.086    &   \cg 5.277  &  \cg \bf 5.456  &   \bf   5.685  &    5.859  & \cg  \bf  5.991 \\  \\[-0.2cm]
\hline
\hline \\[-0.4cm]
Panel C (average PL) & Horizon & 1 & 2 & 3 & 7 & 14 & 21 & 28 \\[0.01mm]
\hline \\[-0.4cm]
& BAR(7) &    -3.926  & \bf  -3.930  &  \bf   -3.998  &   -3.969  &    -4.093  &    \cg -4.085  &  -4.091\\
\multirow{1}{*}{{{\textbf{with Day-of-the-Week}}}} & {sparse BAR(7)}    &     -3.919  &   -3.993  &    -4.069  &    -3.988  &   -4.068  &    -4.130  &    -4.113 \\
 \multirow{1}{*}{{{\textbf{Dummies}}}} &BAR(1) &     -4.113  &    -4.167  &    -4.174  &    -4.128  &    -4.185  &    -4.220  &    -4.285\\
 & BAR(4) (auto.arima) &     -3.987  &    -4.002  &    -4.004  &   -4.059  &   -4.081  &    -4.127  &    -4.202\\
 \hline\\[-0.4cm]
&  {sparse BAR(7)} & -3.954 & -4.056 & -4.082 & -3.973 & -4.097 & -4.105 & -4.152 \\
&  {sparse BAR(7) (Daily Gas)} &  {-3.927} &   -  {4.020} &    {-4.050} &   {\bf  -3.960} &   {-4.084} &    {-4.083} &  {-4.071} \\
\multirow{4}{*}{{{\textbf{with Seasonal}}}} & B-RU-MIDAS (All-IPI + Surveys) & \bf -3.841 & -3.978 & -4.045 & -3.982 & \cg -4.088 & -4.088 & \bf -4.059 \\
\multirow{4}{*}{{{\textbf{Dummies}}}}& B-RU-MIDAS (Surveys) & -3.893 & -3.974 & -4.021 & -4.013 & \cg -4.087 & \cg -4.089 & -4.080 \\
& B-RU-MIDAS (All-IPI) & -3.884 & -3.977 & -4.043 & -4.024 & \cg -4.068 &  \cg \bf -4.057 & -4.108 \\
& B-RU-MIDAS (All-IPI + Surveys + Monthly Oil) & -3.904 & -3.980 & \cg -4.054 & \cg -3.978 &  -4.060 & -4.089 & \cg -4.084 \\
& B-RU-MIDAS (All-IPI + Surveys + Daily Oil) & -3.919 & -3.993 & \cg -4.007 & \cg -4.001 & -4.091 & -4.079 & \cg -4.078 \\
& B-RU-MIDAS (All-IPI + Surveys + Daily Gas) & \cg   -3.913  &  \cg -3.987  &   \cg   -4.048  &  \cg  -3.995  &   \bf   -4.047  &     -4.123  &   \cg   -4.068 \\
\hline
\end{tabular}
\small{
\begin{tablenotes}
\item \textit{Notes:}
 \item[1] Refer to Section \ref{sec_Model} for details on model formulations. B-RU-MIDAS indicates Bayesian RU-MIDAS with Normal-Gamma prior including lags, seasonal dummies and different exogenous variables. All forecasts are produced with recursive estimation of the models.
 \item[2] Bold numbers indicate models with small RMSE and average CRPS; and big average PL across all the models.
\item[3] Gray cells indicate models that belong to the Superior Set of Models delivered by the \textsc{MCS} procedure at confidence level $10\%$.
\end{tablenotes}
}
\end{threeparttable}
\end{adjustbox}
\label{Ger_Forec_AR3_Seas_6models}
\end{table}

\begin{table}[h!]
\caption{\bf Point and density forecasting measures  for Italy.}
\centering
\vspace{-0.1in}
\begin{justify}
\footnotesize{RMSE (Panel A); average CRPS (Panel B) and average Predictive Likelihood (PL) (Panel C) for all the models with day-of-the-Week or seasonal dummies. The estimation and forecasting sample last 7 years each. The forecasting is provided for horizons $h = 1, 2, 3, 7, 14, 21, 28$.}
\end{justify}
\begin{adjustbox}{width=\textwidth,center=\textwidth}
\begin{threeparttable}
\begin{tabular}{c|lccccccc}
\hline \\[-0.4cm]
Panel A (RMSE) & Horizon & 1 & 2 & 3 & 7 & 14 & 21 & 28 \\[0.01mm]
\hline \\[-0.4cm]
 & BAR(7) &     6.661  & 7.012  & \cg  7.271  &  \cg 8.161  &  9.174  &  \cg 9.883  &    10.503\\
 \multirow{1}{*}{{{\textbf{with Day-of-the-Week}}}} & {sparse BAR(7)}  &   7.316  &    7.966  &  8.233  &    8.236  &    9.260  &    10.007  &   10.605\\
 \multirow{1}{*}{{{\textbf{Dummies}}}} & BAR(1) & \cg     8.286 &  8.890 &     9.216 &    9.231 &   10.312 &    10.991 &    11.674 \\
 &  BAR(4) (auto.arima) &     7.299 & \cg 7.448 &   7.280 &    8.652 &    9.699 &  \cg 10.320 &   11.055   \\
\hline \\[-0.4cm]
& {sparse BAR(7)}  & 8.106 & 8.867 & 9.061 & 8.412 & 9.378 & 10.093 & 10.635 \\
& {sparse BAR(7) (Daily Gas)} &      {7.976} &  {8.666} &  {8.832} &   {8.251} & {9.170} &   {9.876} &  {10.376} \\
\multirow{4}{*}{{{\textbf{with Seasonal}}}}& B-RU-MIDAS (All-IPI + Surveys) &\cg  6.618 & \cg 7.030 & \cg 7.289 & \cg 7.962 & \cg 8.861 & 9.434 & \cg 9.969 \\
\multirow{4}{*}{{{\textbf{Dummies}}}}&B-RU-MIDAS (Surveys) & 6.638 & 7.101 & 7.390 & 8.053 & \cg 9.046 & 9.694 & \cg 10.261 \\
& B-RU-MIDAS (All-IPI) & \cg 6.610 & \cg 7.034 & \cg 7.313 & \cg 8.000 & \cg 8.954 & 9.557 & \cg 10.101 \\
& B-RU-MIDAS (All-IPI + Surveys + Monthly Oil) & 6.662 & 7.068 & 7.331 & 7.972 & 8.849 & 9.376 & 9.879 \\
& B-RU-MIDAS (All-IPI + Surveys + Daily Oil) & \bf 6.561 &  \bf 6.956 &  \bf 7.211 &  \bf 7.842 &  \bf 8.686 &  \bf 9.209 &  \bf 9.685 \\
 & B-RU-MIDAS (All-IPI + Surveys + Daily Gas) &    6.569   &  6.970  &    7.229  &  7.879  &  8.786  &   9.363  &   9.896 \\ \\[-0.2cm]
\hline
\hline \\[-0.4cm]
Panel B (average CRPS) & Horizon & 1 & 2 & 3 & 7 & 14 & 21 & 28 \\[0.01mm]
\hline \\[-0.4cm]
 &   BAR(7) &     3.579  &  3.765  & \cg    3.893  &  \cg  4.384  &  \cg    4.920  &     \cg 5.267  &  5.654 \\
 \multirow{1}{*}{{{\textbf{with Day-of-the-Week}}}} &  {sparse BAR(7)}  &     3.936 &    4.297 &     4.427  &     4.433 &  4.970  &      5.328  &  5.702 \\
\multirow{1}{*}{{{\textbf{Dummies}}}} &BAR(1) &     4.456  &    4.798  &     4.968  &  4.961  &     5.575  &     5.887  &    6.263 \\
 &  BAR(4) (auto.arima) &    \cg   3.910 &       3.997 &       3.906 &     4.632 &        5.240 &      \cg   5.527 &        5.931 \\
\hline \\[-0.4cm]
& {sparse BAR(7)} & 4.412 & 4.833 & 4.925 & 4.536 & 5.033 & 5.370 & 5.713 \\
& {sparse BAR(7) (Daily Gas)}  &     {4.331} &     {4.710} &      {4.789} &    {4.459} &     {4.926} &      {5.264} &    {5.578}   \\
\multirow{4}{*}{{{\textbf{with Seasonal}}}}& B-RU-MIDAS (All-IPI + Surveys) & \cg 3.555 & \cg 3.783 & \cg 3.913 & \cg 4.312 & \cg 4.819 & \cg 5.121 & \cg 5.445 \\
\multirow{4}{*}{{{\textbf{Dummies}}}}&B-RU-MIDAS (Surveys) & 3.574 & 3.826 & 3.974 & 4.348 & 4.888 & 5.189 & \cg 5.543 \\
& B-RU-MIDAS (All-IPI) & \cg 3.551 &  3.784 & \cg 3.925 & \cg 4.331 & \cg 4.869 & 5.180 & \cg 5.513 \\
& B-RU-MIDAS (All-IPI + Surveys + Monthly Oil) & 3.555 & 3.772 & 3.902 & 4.279 & 4.760 & 5.038 & 5.338 \\
& B-RU-MIDAS (All-IPI + Surveys + Daily Oil) &  \bf 3.511 &  \bf 3.722 &  \bf 3.846 &  \bf 4.217 &  \bf 4.680 &  \bf 4.950 &  \bf 5.237 \\
  & B-RU-MIDAS (All-IPI + Surveys + Daily Gas) &     3.514  & 3.729  &  3.859  &    4.243  &  4.754  &  5.058  &   5.378\\
\\[-0.2cm]
\hline
\hline \\[-0.4cm]
Panel C (average PL) & Horizon & 1 & 2 & 3 & 7 & 14 & 21 & 28 \\[0.01mm]
\hline \\[-0.4cm]
 &   BAR(7) &       -3.467 &    -3.522 &    -3.573 &   -3.691 &    -3.806 &    \cg -3.871 &    -3.903\\
 \multirow{1}{*}{{{\textbf{with Day-of-the-Week}}}} &  {sparse BAR(7)}  &  -3.546 &    -3.672 &  -3.689 &  -3.684 &    -3.770 &    -3.881 &   -3.896 \\
\multirow{1}{*}{{{\textbf{Dummies}}}} &BAR(1) &     -3.695 &    -3.769 &   -3.833 &    -3.772 &   -3.872 &    -3.962 &   -4.031 \\
 &  BAR(4) (auto.arima) &     -3.604 &    -3.594 &   -3.571 &    -3.756 &    -3.808 &  -3.891 &    -3.972\\
\hline \\[-0.4cm]
& {sparse BAR(7)} & -3.625 & -3.733 & -3.738 & -3.696 & -3.804 & -3.903 & -3.913 \\
& {sparse BAR(7) (Daily Gas)} &  {-3.606}    & {-3.709}    & {-3.739}    & {-3.689}   & {-3.753}   & {-3.869}  & {-3.875} \\
\multirow{4}{*}{{{\textbf{with Seasonal}}}}& B-RU-MIDAS (All-IPI + Surveys) & \cg -3.464 & \cg -3.528 & \cg -3.560 & \cg -3.656 & \cg -3.739 & \cg -3.829 & \cg -3.846 \\
\multirow{4}{*}{{{\textbf{Dummies}}}}& B-RU-MIDAS (Surveys) & \cg -3.464 &  \bf -3.505 & -3.558 & -3.645 & -3.748 & -3.821 & \cg -3.866 \\
&  B-RU-MIDAS (All-IPI) &\cg  -3.447 &\cg  -3.490 & \cg -3.569 & \cg -3.647 &  \cg \bf -3.729 & \cg -3.787 & \cg -3.839 \\
&  B-RU-MIDAS (All-IPI + Surveys + Monthly Oil) & -3.470 & -3.536 & -3.573 & -3.656 & -3.780 & -3.815 & -3.889 \\
&  B-RU-MIDAS (All-IPI + Surveys + Daily Oil) & -3.456 & -3.520 & \cg -3.593 & \cg -3.653 & \cg -3.784 & -3.807 & -3.874 \\
&   B-RU-MIDAS (All-IPI + Surveys + Daily Gas) &   \bf -3.431 &  -3.507 &   \bf -3.555 &   \bf -3.641 &  -3.742 &   \bf   -3.785 &   \bf  -3.821 \\
 \hline
\end{tabular}
\small{
\begin{tablenotes}
\item \textit{See Notes in Table \ref{Ger_Forec_AR3_Seas_6models}}
\end{tablenotes}
}
\end{threeparttable}
\end{adjustbox}
\label{Ita_Forec_AR3_Seas_6models}
\end{table}

\end{document}